\documentclass[%
reprint,
superscriptaddress,
groupedaddress,
 amsmath,amssymb,
aps,
prb,
]{revtex4-1}
\usepackage{graphicx}  % needed for figures
\usepackage{dcolumn}   % needed for some tables
\usepackage{bm}        % for math
\usepackage{verbatim}  % for math
\usepackage{sidecap}

\usepackage{physics}
\usepackage{float}
\usepackage{color}
\usepackage[dvipsnames]{xcolor}
\usepackage[colorlinks]{hyperref}

\def \dag{^{\dagger}}
\def \ham{\mathcal{H}}
\def \S{\mathbf{S}}
\newcommand{\vect}[1]{\mathbf{\bm{#1}}}
\newcommand{\uvec}[1]{\hat{\vect{#1}}}

\makeatletter
\newcommand{\thickhline}{%
    \noalign {\ifnum 0=`}\fi \hrule height 1pt
    \futurelet \reserved@a \@xhline
}
\newcolumntype{"}{@{\hskip\tabcolsep\vrule width 1pt\hskip\tabcolsep}}
\makeatother

\usepackage{cleveref}

\begin{document}

\title{Theory of electronic magnetoelectric coupling in $d^5$ Mott insulators}
\author{Adrien Bolens}
\email{bolens@spin.phys.s.u-tokyo.ac.jp}
%\author{Hosho Katsura}
%\author{Masao Ogata}
%\author{Seiji Miyashita}
\affiliation{
    Department of Physics, University of Tokyo, Hongo, Bunkyo-ku, Tokyo 113-0033, Japan
   }

\date{\today}
\begin{abstract}
Motivated by recent terahertz (THz) spectroscopy measurements in $\alpha$-RuCl$_3$, we develop a theory for magnetoelectric (ME) effects in Mott insulators of $d^5$ transition metal ions in an octahedral crystal field. For $4d$ and $5d$ compounds, the relatively wide-spread orbitals favor charge fluctuations of localized electrons to neighboring ions and a significant ME effect from electronic mechanisms is expected. From a three-orbital Hubbard model with strong spin-orbit coupling, we derive the mechanisms for the electric polarization originating from virtual hopping of the localized holes carrying the spins. We consider the electric polarization generated by pairs of spin operators on nearest neighbor bonds with either an edge-sharing geometry (i.e., two ligands are shared) or a corner-sharing geometry (i.e., one ligand is shared). The allowed couplings are first derived using a symmetry approach. Then, we explicitly calculate the coupling constants and evaluate the effective polarization operator in the ground state manifold using perturbation theory and exact diagonalization. The results are relevant when considering the THz optical conductivity of magnetic systems such as some perovskite iridates or Kitaev materials. In particular, they help explain the recent THz optical measurements of $\alpha$-RuCl$_{3}$ for which the electric-dipole-induced contribution has been shown to be strong.
%
%
%Starting from a symmetry group approach, we build a theory for the ME couplings on nearest neighbor bonds with edge-sharing and corner-sharing geometries of the neighboring octahedral complexes.
%
%involving two-spin operators
%
%The results derived are relevant when considering the terahertz optical conductivity of magnetic systems such as some perovskite iridates or Kitaev materials. In particular, they might help explain recent terahertz optical measurement of $\alpha-$RuCl$_{3}$ for which the electric-dipole-induced contribution has been shown to be strong.
%
%
%Magnetoelectric effects in $d^5$ transition metal Mott insulators surrounded by octahedral ligand complexes are theoretically studied using a multi-band Hubbard model.
%
%We derived the allowed coupling between spins and electric polarization arising purely from the charge of the electrons carrying the spins. Such electronic mechanisms are known in the context of $t_{2g}$ orbital systems in the spin-current model of multiferroics, and for other spin-charge coupling phenomena in Mott insulators originating from virtual hopping of the electrons. % In the presence of inversion symmetry, we point out the importance of spin orbit coupling.
% The results derived are relevant when considering the terahertz optical conductivity of magnetic systems such as some perovskite iridates or Kitaev materials. In particular, they might help explain recent terahertz optical measurement of $\alpha-$RuCl$_{3}$ for which the electric-dipole-induced contribution has been shown to be strong.
 \end{abstract}

\maketitle

\section{Introduction}

Multiferroics are defined by the existence of the static magnetoelectric (ME) effect, i.e., the control of electric polarization by a magnetic field and of magnetization by an electric field \cite{kimura2003magnetic,wang2003epitaxial,hur2004electric,fiebig2005revival,eerenstein2006multiferroic,cheong2007multiferroics,tokura2014multiferroics}. In the dynamical regime, the magnetically induced ferroelectricity leads to the emergence of electromagnons, low-lying modes associated with the hybridization of spin waves with the electric polarization \cite{pimenov2006possible,sushkov2007electromagnons,katsura2007dynamical,cazayous2008possible,seki2010electromagnons,takahashi2012magnetoelectric,chun2018electromagnon}. 

However, ME effects are not limited to multiferroics. In magnetic Mott insulators with lattice inversion invariance and no electric polarization in the ground state, i.e., without the static ME effect and reduction of the symmetry accompanying the long-range ordering, dynamical ME effects are still possible.
%Those new excitations are typically observed through the photo-induced oscillation of the spin-coupled electric polarization. This entanglement of magnetic and electric properties in the dynamical regime shares the same origin as the ME ground state, a mechanism behind the ME coupling. 

%The coupling typically arises from magnetoelastic effects, i.e., from the displacement of some ions. 
Three types of mechanisms for the ME coupling between the local electric polarization $\vb P_{ij}$ and spin operators at sites $i$ and $j$ are typically considered for multiferroics: (i) the exchange-striction mechanism arising from the symmetric spin exchange interaction $\vb S_i \cdot \vb S_j$ \cite{sergienko2006ferroelectricity}, (ii) the spin-current model arising from the antisymmetric spin exchange interaction $\vb S_i \times \vb S_j$ (or inverse Dzyaloshinskii-Moriya interaction) \cite{katsura2005spin,mostovoy2006ferroelectricity,sergienko2006role}, and (iii) the spin dependent $p$-$d$ hybridization mechanism, which causes single spin anisotropy \cite{jia2006bond,jia2007microscopic,arima2007ferroelectricity}. Only the mechanism (ii) is allowed in a system with inversion symmetry, whereas the lack of inversion symmetry centered at the middle of bonds and spin sites is necessary for mechanisms (i) and (iii), respectively.

The ME coupling affects the charge dynamics of the electrons in the magnetic energy scale, far below the optical gap of Mott insulators. Thanks to the coupling, the magnetic excitations respond to an AC electric field and can be probed, for example, by measuring the terahertz (THz) dielectric response. This is especially interesting for quantum spin liquids (QSL) with fractionalized excitations, which result in a continuous subgap optical conductivity. This effect has been predicted and observed in some QSLs \cite{ng2007power, potter2013mechanisms, elsasser2012power, pilon2013spin}, and explained with a different kind of ME mechanism. For the single-band Hubbard model, charge effects are possible through virtual hopping of the electrons, which scales as $(t/U)^n$ for loops of size $n$ (with $t$ the hopping amplitude and $U$ the on-site repulsion) \cite{bulaevskii2008electronic}. Even without spin-orbit coupling (SOC), a nontrivial ME coupling arises for odd $n$, starting from triangular loops, and explains the subgap optical conductivity of gapless QSLs on triangular and kagome lattices \cite{ng2007power,potter2013mechanisms, huh2013optical, hwang2014signatures}.

More recently, a subgap optical conductivity was observed in the Kitaev material $\alpha$-RuCl$_3$ in a series of THz spectroscopy measurements \cite{little2017antiferromagnetic, wang2017magnetic, wellm2017signatures, reschke2018sub, shi2018field}. In a previous work, we showed that the virtual hopping mechanism can be relevant for $n=2$ (single bonds) when considering multiorbital systems such as Kitaev materials with strong SOC \cite{bolens2018mechanism}.

In this paper, motivated by the THz spectroscopy results in $\alpha$-RuCl$_3$, we derive the general mechanisms for the polarization operators of $4d^5$ and $5d^5$ Mott insulators arising from the electronic charge in $t_{2g}$ orbitals.

 More specifically, we consider systems with inversion symmetry and without degeneracy of the local ground states (other than the Kramers degeneracy), so that the low-energy manifold is described with pseudospin-1/2 variables on each site. We focus on systems with $d^5$ transition metal (TM) ions, each surrounded by an octahedral complex of ligands that connect neighboring TM sites. For $4d$ and $5d$ compounds, the wide-spread orbitals favor charge fluctuations of the localized holes to neighboring ions, so that we expect significant ME effects from purely electronic mechanisms.
  
\begin{figure}
\centering
	\includegraphics[width=0.42\textwidth]{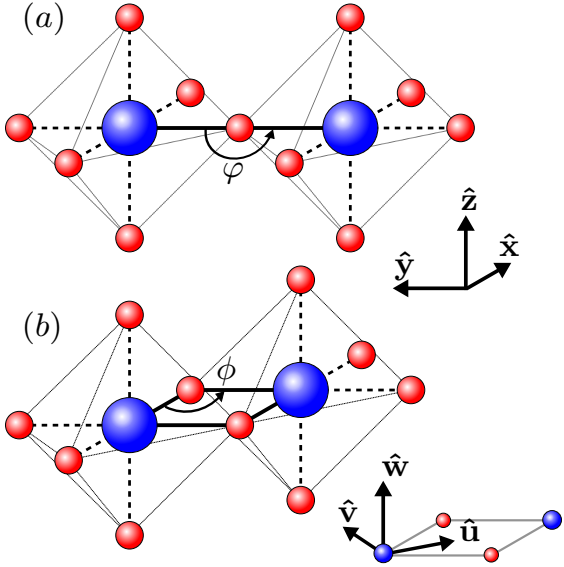}
	\caption{(a) Corner-sharing and (b) edge-sharing geometries. The blue spheres represent the TM ions and the smaller red spheres the ligands. Deviations from $\phi = 90^{\circ}$ and $\varphi = 180^{\circ}$ are considered in Appendices~\ref{sec:edge_angle} and \ref{sec:corner_angle}.}
\label{fig:geometries}
\end{figure}

  Two geometries are considered, the edge-sharing geometry (i.e., two ligands are shared) and the corner-sharing geometry (i.e., one ligand is shared) as shown in Fig.~\ref{fig:geometries}.

%For $4d$ and $5d$  compounds, 
%The dynamical ME effects can be explained through purely electronic mechanisms which only rely on the wavefunctions. This is especially expected in Mott insulators with outspread orbitals which favors the charge fluctuation of localized electrons to neighboring ions, such as some $4d$ and $5d$ transition metal (TM) compounds.
%

%
%However, ME effects are not limited to multiferroics. In magnetic materials with lattice inversion invariance and no electric polarization in the ground state, i.e., without the static ME effect and reduction of the symmetry accompanying the long-range ordering, dynamical ME effects are still possible.
%%
%such that the electric polarization vanishes in the ground state and there is no static ME effects. Nevertheless, dynamical ME effects are still possible.

%  The magnons of magnetically ordered systems can in principle also be probed with the AC electric field through the ME coupling. The polarization is odd under timer-reversal symmetry hence the coupling only involves terms even in spin operators. The subgap weight thus originates form multi-magnon excitations and is continuous.
 
% While magnetoelatstic mechanisms are not excluded, 
 
%  of the electrons an not the displacement of the ions.

Our goal is to calculate the effective polarization in the magnetic ground state manifold, i.e., in terms of spin-1/2 operators. Due to time-reversal symmetry, there are no one-spin terms and we only consider terms written with two-spin operators from nearest neighbors.

 The spin-current model, which lead to the Katsura-Nagaosa-Balatsky (KNB) formula $\vb P_{ij} \propto \vu e_{ij} \times (\vb S_i \times \vb S_j)$ \cite{katsura2005spin}, where $\vu e_{ij}$ is the unit vector parallel to the bond, was originally based on a purely electronic mechanism made possible by finite matrix elements such as $\mel{d_{xy}}{y}{p_x}$ between neighboring TM ions and ligands \cite{katsura2005spin}. The same matrix elements have been shown to give rise to additional symmetry-allowed couplings between $\vb P_{ij}$ and $\vb S_i \times \vb S_j$ when the symmetry is low enough \cite{kaplan2011canted, miyahara2016theory, matsumoto2017symmetry}.
%  Although responsible for a finite polarization in multiferroics when the inversion symmetry is spontaneously broken by the long-range ordering, it is also relevant in the dynamical regime of systems with no finite polarization in the ground state.

%Another type of mechanism behind the ME coupling originates from the charge dynamics in the Hubbard model and do not require different orbitals. For the single-band Hubbard model, charge effects are only possible through virtual hopping of the electrons, which scales as $(t/U)^n$ for loops of size $n$ (with $t$ the hopping amplitude and $U$ the on-site repulsion). In the single-band case, even without spin orbit coupling (SOC), a non-trivial coupling arises for odd $n$, starting from triangular loops, and explains the sub-gap optical conductivity of gapless spin liquids on triangular and Kagome lattices. In a previous work, we showed that the same virtual hopping mechanism is relevant for $n=2$ (i.e. single bonds) when considering multi orbital systems such as Kitaev materials.

In short, the position operator $\vb r$ has two types of finite matrix elements in the tight-binding formalism. The KNB type $\mel{i, \alpha}{\vb r}{j, \beta}$ for ions $i\neq j$ (typically one TM ion and one ligand) with orbitals $\alpha$ and $\beta$, and the lattice type $\mel{i, \alpha}{\vb r}{i, \beta} = \vb r_i \delta_{\alpha, \beta}$ where $\vb r_i$ is the position of ion $i$. If combined with first-principle calculations to accurately evaluate the  $\mel{i, \alpha}{\vb r}{j, \beta}$ integrals, the ME effect in Kitaev materials and other $d^5$ Mott insulators can be evaluated from our results.

This paper is organized as follows. First, in Sec.~\ref{sec:model}, we introduce the Hamiltonians. In Sec.~\ref{sec:symmetry}, we derive the allowed ME couplings using a symmetry group approach. In addition, we stress the importance of SOC in Sec.~\ref{sec:soc}. In Sec.~\ref{sec:polarization}, we define two different electronic mechanisms for the polarization starting from the usual LCAO (linear combination of atomic orbitals) approximation of the tight-binding model. We then explicitly consider the mechanisms in Sec.~\ref{sec:mechanism} in the edge- and corner-sharing geometries for large SOC. Finally, we discuss our findings and conclude in Sec.~\ref{sec:discussion}.

%We consider first an approach based on group theory to . Then we investigate in details the different mechanism possible. We consider a fixed lattice, and thus only the polarization originating from the electron wave-functions. 
%
%Starting from the usual LCAO (linear combination of atomic orbitals) of the tight-binding model, we derive two different mechanisms. First, the "lattice" contribution arises from the virtual hopping of the electron on neighboring sites. It simply depends on the positions of the electrons on the lattice, which is why we call it the "lattice-polarization". Secondly, the "orbital" contribution arises from the integrals of the polarization operator and the atomic orbital wave-functions. Because only wave-function on different sites give a finite integral, the "orbital-polarization" behave like a hopping operator once casted on the tight-binding formalism. Both those mechanism are carefully defined and investigated in the following.

\section{Model}
\label{sec:model}
We consider Mott insulators with unfilled $t_{2g}$ orbitals under an octahedral crystal field (CF) of chalcogens or halogens, which strongly split the $t_{2g}$ and $e_g$ orbitals. In the $d^5$ configuration, there is one hole in the $t_{2g}$ orbitals per site. Accordingly, we consider a three-band Hubbard model for the holes on the lattice of TM ions, with a filling corresponding to one hole per site,
\begin{equation}
\label{eq:ham_general}
  \ham = \ham_{\textrm{hop}} + \ham_{\textrm{ion}} + \ham_{\textrm{int}},
\end{equation}
which is the sum of a hopping, on-site, and interaction Hamiltonian, respectively. Because we are interested in bonds made of two neighboring TM ions, it is enough to consider only two sites.
The Hamiltonians are concisely expressed by using the hole operators
\begin{equation}
  \mathbf{c}_i \dag=(c\dag_{i, yz,\uparrow},c\dag_{i, yz,\downarrow},c\dag_{i, xz,\uparrow},c\dag_{i, xz,\downarrow},c\dag_{i, xy,\uparrow},c\dag_{i, xy,\downarrow}).
\end{equation}
States with doubly occupied sites (thereafter called polar states) are affected by the Coulomb interaction and Hund's coupling as described by the Kanamori Hamiltonian \cite{kanamori1963electron,georges2013strong}
%\cite{kanamori1963electron,georges2013strong}
with intra orbital Coulomb repulsion $U$, inter-orbital repulsion $U'=U - 2J_H$, and Hund's coupling $J_H$,
\begin{align}
\label{eq:kanamori}
  \ham_{\rm int} = &U \sum_{i,a} n_{i,a,\uparrow} n_{i,a,\downarrow} + (U' -J_H) \sum_{i, a<b, \sigma} n_{i,a,\sigma}n_{i,b,\sigma} \nonumber \\
  & + U'_{i,a\neq b} n_{i,a,\uparrow} n_{i,b,\downarrow} - J_H\sum_{i,a\neq b} c_{i,a,\uparrow} \dag c_{i,a,\downarrow} c_{i,b,\downarrow}\dag c_{i,b,\uparrow} \nonumber \\
  & + J_H \sum_{i,a\neq b} c_{i,a,\uparrow}\dag c_{i,a,\downarrow}\dag c_{i,b,\downarrow} c_{i,b,\uparrow}.
\end{align}

Each hole can jump onto the filled $p^6$ orbitals of the ligands at a cost $\Delta_{pd}$, the charge-transfer energy.
%(for the hole, this also includes the Coulomb energy needed to leave the TM ion)
Additionally, direct hopping between TM ions are also possible. 
We consider the limit where $\Delta_{pd} \gg U$, i.e., the system is in a Mott insulating phase (in contrast with the charge-transfer insulator), so that intermediate states with unfilled $p$ orbitals can be integrated out to give the effective Hamiltonian~\eqref{eq:ham_general}. Moreover, in this limit, intermediate states with two or more unfilled $p$ orbitals are neglected and the effective hopping integrals are simply obtained by considering the different TM-L-TM hopping processes through each ligand (L) separately. Summing the contributions from all ligands, we obtain 
\begin{equation}
  \ham_{\rm hop} = -\sum_{ \langle ij \rangle } \mathbf{c}_i\dag ( \hat{T}_{ij} \otimes \mathbb{I}_{2\times 2} ) \mathbf{c}_j,
\end{equation}
where $\hat{T}_{ij}$ is the hopping matrix between sites $i$ and $j$,
\begin{equation}
  T_{ij}^{\alpha \beta} =  t_{ij}^{\alpha \beta} + \sum_{p} \sum_{\gamma=x,y,z}\frac{t_{i p}^{\alpha \gamma} t_{jp}^{\beta \gamma}} {\Delta_{pd}}.
\end{equation}
Here $t_{ij}^{\alpha \beta}$ is the hopping between the two TM ions at sites $i$ and $j$ with respective orbitals $\alpha$ and $\beta$, $p$ labels the ligands between the two TM ions, $\gamma$ labels their orbitals, and $t_{ip}^{\alpha \gamma}$ is the hopping between the TM ion $i$ and ligand $p$ with orbitals $\alpha$ and $\gamma$, respectively. The different hopping amplitudes are the usual Slater-Koster integrals \cite{slater1954simplified}. 

Finally, the on-site six-dimensional Hilbert space is further split by $\ham_{\rm ion}$. Each hole has an effective $L=1$ orbital angular momentum in addition to its spin-$1/2$ angular momentum.
The on-site Hamiltonian consists of SOC,
\begin{equation}
  \ham_{\textrm{SOC}} = \frac{\lambda}{2} \sum_{i,a} {\bf c}^\dagger_i (L^a \otimes \sigma^a) {\bf c}_i
\end{equation}
with $\lambda >0$, 
and additional CF splitting,
\begin{equation}
  \ham_{\rm CF} = \Delta \sum_i \mathbf{c}_i\dag [ \qty(\mathbf{L}\cdot \uvec{n}_{\rm CF})^2 \otimes \mathbb{I}_{2\times 2}]\mathbf{c}_i,
\end{equation}
so that $\ham_{\rm ion} =  \ham_{\rm SOC} +  \ham_{\rm CF}$.
Here, $(L^a)_{bc}= -i \epsilon_{abc}$ and $\sigma^a$ are the Pauli matrices. The unit vector $\uvec{n}_{\rm CF}$ depends on the system considered. In this paper, only the trigonal ($\uvec{n}_{\rm CF}=[111]$) and tetragonal ($ \uvec{n}_{\rm CF}=[001]$) distortions are explicitly considered in the edge-sharing and corner-sharing geometries, respectively. They correspond (up to a constant) to
\begin{equation}
  \label{eq:CF_matrices}
   \mathbf{L}_{[111]}^2 =
  -\frac{1}{3}\begin{pmatrix}
    0 & 1 & 1 \\
	1 & 0 & 1 \\
	1 & 1 & 0
  \end{pmatrix} \textrm{ and }
\mathbf{L}_{[001]}^2 = 
  \begin{pmatrix}
    1 & 0 & 0 \\
	0 & 1 & 0 \\
	0 & 0 & 0
  \end{pmatrix}.
\end{equation}
The SOC Hamiltonian splits the $t_{2g}$ orbitals into states with total effective angular momentum $J=1/2$ and $J=3/2$ with energies $-\lambda$ and $\lambda/2$, respectively. The CF distortion by itself also splits the three $t_{2g}$ states into a $b_{2g}$ state and two $e_{g}$ states. Generally, the combination of SOC and CF splits the states into three Kramers doublets.

We consider the situation where pseudospin-1/2 variables can safely be defined. This implies that the ground state is a doublet (always the case with $\lambda >0$) that is sufficiently gapped from the other four states; it should be larger than the magnetic scale $t_{\textrm{eff}}^2/U$. (Here $t_{\textrm{eff}}$ refers generally to the amplitude of the effective hopping between neighboring TM sites.)
In the next section, we consider the allowed ME couplings on a symmetry basis, and do not make additional assumptions for the on-site Hamiltonian.
 In Sec.~\ref{sec:mechanism}, when explicitly considering the microscopic mechanisms, we will make the assumption that SOC is large, $\lambda \gg t_{\textrm{eff}}^2/U$. Moreover, the additional CF distortion, whose main effect is to lower the symmetry, will be introduced perturbatively in analytical calculations. This situation is particularily relevant for late TM ions such as Ir, Os, Rh, and Ru \cite{jackeli2009mott}.
 
\begin{figure}
\centering
	\includegraphics[width=0.3	\textwidth]{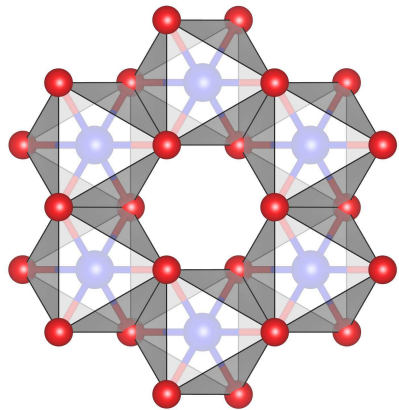}
	\caption{Crystal structure of Kitaev materials. The blue spheres are the TM ions (Ir or Ru) and the red spheres are the ligand (O or Cl). Each TM-TM bond is in the edge-sharing geometry.}
\label{fig:honeycomb}
\end{figure}

 In this limit, the present model on the two-dimensional honeycomb lattice (with edge-sharing bonds, see Fig.~\ref{fig:honeycomb}) has been discussed extensively \cite{rau2014generic,sizyuk2014importance,rau2014trigonal,winter2016challenges,kim2016crystal,yadav2016kitaev,winter2017models,hou2017unveiling} in the context of Kitaev materials such as $\alpha$-RuCl$_3$ \cite{plumb2014alpha,banerjee2016proximate}and `Kitaev iridates' A$_2$IrO$_3$ with A=Na, Li \cite{singh2010antiferromagnetic,singh2012relevance}. The model on the square lattice (with corner-sharing bonds) has been discussed in the context of iridates or rhodates compounds such as Sr$_2$IrO$_4$ and Sr$_2$Ir$_{1-x}$Rh$_x$O$_4$ \cite{perkins2014interplay}.

\section{Symmetry considerations}
\label{sec:symmetry}

Group theory is a powerful tool to derive the allowed spin-polarization coupling \cite{kaplan2011canted, miyahara2016theory, matsumoto2017symmetry}.
Here we explicitly consider the two geometries depicted in Fig.~\ref{fig:geometries} with and without the inclusion of CF distortions.

In the ground state manifold (one pseudospin 1/2 per TM ion), because of the time-reversal and inversion symmetry, the general form of the coupling between the electric polarization and the pseudospins (referred to as simply spins from now on) for a given bond $\langle ij \rangle$ 
is $\vb P = \hat m \qty(\vb S_i \times \vb S_j)$, or
\begin{equation}
\label{eq:coupling_withinv}
  \begin{pmatrix}
  	P_x \\
  	P_y \\
  	P_z
  \end{pmatrix}
   = 
  \begin{pmatrix}
    m_{xx} & m_{xy} & m_{xz} \\
	m_{yx} & m_{yy} & m_{yz} \\
	m_{zx} & m_{zy} & m_{zz}
  \end{pmatrix}
  \begin{pmatrix}
  \qty(\vb S_i \times \vb S_j)_x \\ 	
  \qty(\vb S_i \times \vb S_j)_y \\ 	
  \qty(\vb S_i \times \vb S_j)_z \\ 	
  \end{pmatrix}.
\end{equation}
 
For a bond in the $x$ direction, the KNB coupling corresponds to the case $m_{yz} = - m_{zy}$ and all other components equal to zero. In general, the allowed matrix elements of $\hat m$ are dictated by the symmetry group of the bond $\langle ij \rangle$. 

\subsection{Edge-sharing geometry}
\label{sec:edge}

Let us first consider the situation where each TM ion is surrounded by six ligands forming edge-sharing octahedra: Neighboring TM ions share two ligands each forming a $\phi = 90^{\circ}$ TM-L-TM bond as depicted in Fig.~\ref{fig:geometries}(b).

\subsubsection{Full octahedral symmetry}
Ideally, the ligands form a perfect octahedra around each TM ion in the $\pm \vu x$, $\pm \vu y$ and $\pm \vu z$ directions, in which case the point group of the system is $O_h$. 
%The $e_g$ and $t_{2g}$ irreducible representation for the $d$ orbitals are split. 
Let us consider a bond in the in the $(\vu x - \vu y)/\sqrt{2}$ direction (also called a $Z$ bond).
The other bonds in the honeycomb lattice of Fig.~\ref{fig:honeycomb} are simply related by cyclic permutations.

The structure has three $C_2$ symmetries along the $\vu u = (\vu x - \vu y)/\sqrt{2}$, $\vu v = (\vu x + \vu y)/\sqrt{2}$ and $\vu w = \vu z$ axis. The symmetry group is thus $D_{2h}$. The character table $D_{2h}$ is shown in Table \ref{tab:D2} of Appendix \ref{sec:tables} along with the character tables of other relevant groups. 

The $T_2$ representation of $O$ breaks into three different irreducible representations of $D_{2h}$: $T_2 \rightarrow B_1 \oplus B_2 \oplus B_3$ for the three $d_{ab}$ orbitals where $a$ and $b$ correspond to the $\vu u$, $\vu v$, and $\vu w$ directions.
%Similarly, for the $e_g$ orbitals, $E \rightarrow A \oplus A$.
The effective hopping (after integrating out the ligand) between the $t_{2g}$ orbitals of sites $i$ and $j$ therefore must be
 \begin{align}
  \hat T_Z(D_{2h}) = 
  \begin{pmatrix}
    t_{11} & 0 & 0 \\
	0 & t_{22} & 0 \\
	0 & 0 & t_{33}
  \end{pmatrix}_{\mathcal{B}',\mathcal{B}'}
  =
  \begin{pmatrix}
    t_{1} & t_2 & 0 \\
	t_2 & t_{1} & 0 \\
	0 & 0 & t_{3}
  \end{pmatrix}_{\mathcal{B},\mathcal{B}},
\end{align}
where $\mathcal{B}$ and $\mathcal{B}'$ are the $\{ \vu x, \vu y, \vu z \}$ and $\{ \vu u, \vu v, \vu w \}$ bases, respectively. 
%We note that to perform the rotations from $\mathcal{B}'$ to $\mathcal{B}$, the $e_{g}$ orbitals also have to be considered.
The hopping integrals can be expressed via the Slater-Koster integrals \cite{slater1954simplified, winter2016challenges} ($t_{pd\sigma}, t_{pd\pi}, t_{dd\sigma}, t_{dd\pi}$ and $t_{dd\delta}$),
\begin{align}
t_1 &= \frac{t_{dd\pi} + t_{dd\delta}}{2}, \nonumber \\ 
t_2 &= \frac{- t_{dd\pi} + t_{dd\delta}}{2} + \frac{t_{pd\pi}^2}{\Delta_{pd}}, \nonumber \\
t_3 &= \frac{3t_{dd\sigma} + t_{dd\delta}}{4}.
\end{align}

The spin operators $\vb S \cdot \vu u$, $\vb S \cdot \vu v$, and $\vb S \cdot \vu w$ transform as $B_{1,g}$, $B_{2,g}$, and $B_{3,g}$, respectively, and the antisymmetrization between site $i$ and $j$ from the cross product acts as an extra $B_{1,u}$. The polarization $P_{\alpha}$ simply acts as a vector: $B_{\alpha, u}$. Hence, only two independent coupling constants are allowed,
\begin{align}
\label{eq:pol_allowed_noCF}
  \hat m(D_{2h}) = 
  \begin{pmatrix}
    0 & 0 & 0 \\
	0 & 0 & m_4 \\
	0 & m_5 & 0
  \end{pmatrix}_{\mathcal{B}',\mathcal{B}'}.
\end{align}
Note that unlike the hopping matrix, which is symmetric, the spin-polarization coupling matrix does not have to be symmetric nor antisymmetric. Already, this is more general than the $\vb P \propto \vu e_{ij} \times (\vb S_i \times \vb S_j)$ coupling, as $m_4$ and $m_5$ are \textit{a priori} not related.
\subsubsection{With trigonal distortion}
Real materials have additional distortions away from the octahedral CF. For Kitaev materials, the most commonly considered distortion is a uniform elongation or compression perpendicular to the plane containing the honeycomb lattice, along the $(\vu x + \vu y + \vu z)/\sqrt{3}$ axis. In this case, the $C_3$ symmetry is intact and the point group of the system is $D_{3d}$. The symmetry group of the bond is then reduced to $D_{2h} \rightarrow C_{2h}$. Its character table is shown in Table \ref{tab:C2} of Appendix \ref{sec:tables}.

In this case, the hopping matrix has one additional allowed term,
 \begin{align}
  \hat T_Z(C_{2h}) = 
  \begin{pmatrix}
    t_{11} & 0 & 0 \\
	0 & t_{22} & t_{23} \\
	0 & t_{23} & t_{33}
  \end{pmatrix}_{\mathcal{B}',\mathcal{B}'}
  =
  \begin{pmatrix}
    t_{1} & t_2 & t_4 \\
	t_2 & t_{1} & t_4 \\
	t_4 & t_4 & t_{3}
  \end{pmatrix}_{\mathcal{B},\mathcal{B}},
\end{align}
Similarly, in the $C_{2h}$ symmetry there are five allowed coupling constants between the spin antisymmetric vector and the polarization:
\begin{align}
\label{eq:pol_allowed_CF}
  \hat m(C_{2h}) = 
  \begin{pmatrix}
    m_1 & 0 & 0 \\
	0 & m_2 & m_4 \\
	0 & m_5 & m_3
  \end{pmatrix}_{\mathcal{B}',\mathcal{B}'}.
\end{align}

\subsubsection{Additional distortions}
In addition to the trigonal distortion, a monoclinic distortion is also usually observed, albeit small. It further lowers the symmetry group of the bond, $C_{2h} \rightarrow C_i$. In this a case, there are no constraints; the hopping matrix has six independent terms and the matrix $\hat m$ has nine. However we will not consider this symmetry situation in the following, and limit ourselves to the trigonal distortion.

A particular distortion is worth mentioning: when, for each bond, the two ligands move toward or away from the center of the bond, in the $\pm \vu v$ direction. In principle, the bond has a $C_{2h}$ symmetry in this case. However, the microscopic processes we consider are derived in the subsystem including only the two TM ions and two ligands. Because the displacement of the ligands is along the $\vu v$ axis, one of the $C_{2}$ axes, the processes will only yield coupling consistent with the $D_{2h}$ symmetry group. In a sense, this is the most practical way to study the TM-L-TM angle dependence away from the ideal $\phi=90^{\circ}$ bond geometry. This case is explicitly considered in Appendix \ref{sec:edge_angle}.

% (I do not know how they introduced $\phi$ in their calculation, but I get the same result for $t_1$ and $t_2$ but not $t_3$).

\subsection{Corner-sharing geometry}
\label{sec:corner}
In the corner-sharing geometry, the octahedral complexes of two neighboring TM ions share one ligand. When an angle is introduced away from $\varphi=180^{\circ}$, the bond inversion symmetry is broken and additional couplings are possible, deviating from Eq.~\eqref{eq:coupling_withinv}. The $\varphi \neq 180$ geometry is considered in Appendix \ref{sec:corner_angle}.

\subsubsection{Full octahedral symmetry}
%The ligands form an octahedra around each TM ion in the $\pm \vu x$, $\pm \vu y$ and $\pm \vu z$ directions.
A bond consists of two octahedra connected along, say, the $\vu x$ direction as depicted in Fig.~\ref{fig:geometries}(a). The structure has four $C_2$ axes, in addition to one $C_4$ axis and inversion symmetry. The symmetry group is thus $D_{4h}$
%(or effectively $D_{\infty h}$ for the hopping processes for which only the shared ligand matters, but this does not affect the outcome).
and its character table is shown in Table \ref{tab:D4}. 
The $T_2$ representation of $O$ breaks into three different irreducible representations of $D_{4h}$: $T_2 \rightarrow B_2 \oplus E$ for the three $t_{2g}$ orbitals. The $d_{yz}$ orbital transforms as $B_{2g}$, while $(d_{xy}, d_{xz})$ transforms as $E_g$.
% Similarly, for the $e_g$ orbitals, $E \rightarrow A \oplus A$.
  The effective hopping (after integrating out the ligand) between the $t_{2g}$ orbitals of sites $i$ and $j$ therefore must be
 \begin{align}
  \hat T_x(D_{4h}) = 
  \begin{pmatrix}
    t_{1} & 0 & 0 \\
	0 & t_{2} & 0 \\
	0 & 0 & t_{2}
  \end{pmatrix}_{\mathcal{B},\mathcal{B}}.
%   \qq{and}
%\hat T_y(D_{4h}) = 
%  \begin{pmatrix}
%    t_{2} & 0 & 0 \\
%	0 & t_{1} & 0 \\
%	0 & 0 & t_{2}
%  \end{pmatrix}.
\end{align}
Neglecting the $d$-$d$ direct hopping, which is safe in the present geometry, we have $t_1=0$ and $t_2=-\frac{t_{pd\pi}^2}{\Delta_{pd}} \equiv t$.
%$t_1=t_{dd\delta}$ and $t_2=-\frac{t_{pd\pi}^2}{\Delta_{pd}} + t_{dd\pi}$

The ME coupling is
\begin{align}
\label{eq:polcorner_allowed_noCF}
  \hat m_x(D_{4h}) = 
  \begin{pmatrix}
    0 & 0 & 0 \\
	0 & 0 & m_1 \\
	0 & -m_1 & 0
  \end{pmatrix}_{\mathcal{B},\mathcal{B}}.
%  \textrm{ and }
%\hat m_y(D_{4h}) = 
%  \begin{pmatrix}
%    0 & 0 & -p^A_1 \\
%	0 & 0 & 0 \\
%	p^A_1 & 0 & 0
%  \end{pmatrix}.
\end{align}
which corresponds to the KNB formula because of the high symmetry of the system (in contrast to the edge-sharing geometry).

\subsubsection{Tetragonal distortion}
A uniform tetragonal distortion only affects the ligands above and bellow the TM ions. It reduces the bond symmetry from $D_{4h}$ to $D_{2h}$. Equation~\eqref{eq:polcorner_allowed_noCF} thus becomes similar to Eq.~\eqref{eq:pol_allowed_noCF} and allows a deviation from the KNB form,
\begin{align}
\label{eq:polcorner_allowed_CF}
  \hat m_x(D_{2h}) = 
  \begin{pmatrix}
    0 & 0 & 0 \\
	0 & 0 & m_1 \\
	0 & m_2 & 0
  \end{pmatrix}_{\mathcal{B},\mathcal{B}}.
%   \textrm{ and }
%\hat m_y(D_{2h}) = 
%  \begin{pmatrix}
%    0 & 0 & -p^A_1 \\
%	0 & 0 & 0 \\
%	-p^A_2 & 0 & 0
%  \end{pmatrix}.
\end{align}
The hopping processes, however, are not affected by the ligand in the $\pm \vu z$ direction and thus by the tetragonal distortion.

%In the case of staggered CF distortion the bond symmetry is $C_{2v}$ (with main axis along the bond). In this case, the bond inversion symmetry is broken even for $\varphi =180$, and there are additional allowed couplings compared to the uniform CF situation described by Eq.~\eqref{eq:coupling_withoutinv} in Appendix~\ref{sec:corner_angle}. In the following, we only explicitly consider the uniform CF.
%
%The distortion affects the ground state manifold and will be relevant when explicitly calculating the matrix elements using exact diagonalization or perturbation theory.

\section{Importance of spin-orbit coupling}
\label{sec:soc}
There are two potential origins for the coupling between spin and charge (or polarization) of the electrons.

The intrinsic Pauli exclusion principle prevents two electrons with the same magnetic state to be on the same site. Alternatively, SOC explicitly couples the motion of an electron with its spin.
Here, we show that SOC is actually essential in order to have a finite electronic ME coupling involving pairs of spin-1/2 operators in systems with inversion symmetry.
% This implies that a strong ME coupling in inversion symmetric systems is expected in relativistic TM materials. 

Without SOC, the orbital degeneracy is only lifted by the CF distortion and the pseudospins of the ground state manifold are the original spins of the holes. Due to inversion symmetry, the polarization operator $\vb P = \hat m(\vb S_i \times \vb S_j)$ only has finite matrix elements between the singlet and triplets in spin space. The microscopic operator $\vb P$ is obviously trivial in terms of electronic spin operators. Therefore, the effective polarization operator in the ground state manifold vanishes without SOC. This argument is only valid when the symmetry group of the bond contains inversion symmetry, without which spin-symmetric matrix elements (i.e., triplet to triplet or singlet to singlet) are possible.

To understand better the role of SOC, we consider the localized single-hole wave functions in the ground state. If we now consider a single bond $\langle ij \rangle$, each wave function transforms as a representation, say $\Gamma$, of the bond symmetry group. Without SOC, $\Gamma$ is one-dimensional with an extra spin degeneracy. With SOC, however, $\Gamma$ is a two-dimensional representation of the corresponding spin double group that mixes orbitals and spins.

Because the wave functions are on different sites, there is an extra site index. For two sites, this extra degree of freedom transforms as one of two different one-dimensional representations of the bond symmetry group so that the two-hole state is overall odd under exchange (as imposed by the fermionic statistic): either $\Gamma_{\text{sym}}=1$, the trivial representation, or $\Gamma_{\text{antisym}}$. The explicit identification of $\Gamma_{\text{antisym}}$ depends on the group (see, e.g., the tables in Appendix~\ref{sec:tables}, where it is indicated as ``i- j antisym''), but it is always odd under inversion symmetry. 

Without SOC, $\Gamma \otimes \Gamma = 1$ and, due to the extra spin degeneracy (singlet or triplet), the full two-hole ground state manifold transforms as $\Gamma_{\textrm{sym}} \oplus \Gamma_{\textrm{antisym}}$. Because the polarization is odd under inversion, only matrix elements between spin triplet and spin singlet are possible when SOC is absent, which, as already mentioned, cannot be finite. Once SOC is introduced, each irreducible representation of $\Gamma \otimes \Gamma = \sum_k \Gamma_k$ is either even or odd under permutation of the single-hole wave functions, and has to be combined accordingly with $\Gamma_{\textrm{sym}}$ or $\Gamma_{\textrm{antsym}}$. However, the dichotomy does not correspond to spin triplet and spin singlet anymore, and the above argument cannot be used.

\section{Polarization}
\label{sec:polarization}
We start by defining the polarization operator $\vb P$ in second quantized notation. This is done by explicitly calculating its matrix elements between the different single-electron wave functions. In this section, the spin degree of freedom is not important and not written explicitly. The Bloch wave functions can be expressed as
\begin{equation}
  \psi_m(\vb{k}, \vb{r}) = \frac{1}{N}\sum_n a_m (\vb{R}_n, r)e^{i\vb k \cdot \vb r},
\end{equation}
where $\vb{R}_n$ are lattice vectors, $N$ is the number of unit cells and $a_m (\vb{R}_n, r)$ is a Wannier function for the unit cells at $\vb{R}_n$. In the tight-binding approach, the LCAO approximation is used for which the exact Wannier functions are replaced by isolated atomic orbitals (or a linear combination of them, when there are multiple atoms per unit cell).

We now consider a single TM-TM bond consisting of two TM ions at positions $\vb{r}^{\rm TM}_1$ and $\vb{r}^{\rm TM}_2$ and $M$ ligands at positions $\vb{r}^{\rm L}_p$ ($p=1,\dots,M$). There are thus $10+3M$ different atomic orbitals centered at $2+M$ different positions,
\begin{equation}
  \psi_{\text{at}}(\vb r) = \left\{
                \begin{array}{ll}
                 \psi_{\alpha}(\vb r - \vb r^{\rm TM}_1), \quad \psi_{\alpha}(\vb r - \vb r^{\rm TM}_2) \qq{($d$ orbitals)}\\
                 \psi_{\gamma}(\vb r - \vb r^{\rm L}_p)\quad p=1,\dots,M  \qq{($p$ orbitals)}
                \end{array}
              \right.
              ,
\end{equation}
where $\alpha \in \{yz,xz,xy,x^2-y^2, 3z^2-r^2 \}$, and $\gamma \in \{x,y, z\}$. We furthermore assume, for simplicity, that the atomic orbitals are orthogonal to each other. The different atomic orbitals are written as $\ket{\vb R, \alpha}$, where $\vb R$ is the position of the corresponding ion and $\alpha$ labels the orbital.
The polarization operators $\vb P= e\vb r$ is then obtained by calculating the different matrix elements $e\mel{\vb R, \alpha}{\vb{r}}{\vb R', \beta}$. In the following, we set the elementary charge $e=1$. In general, $\vb P$ can be decomposed into single-site terms ($\vb R = \vb R'$) and two-site terms ($\vb R \neq \vb R'$),
% (PRL 101 217201) 
 which we call ``lattice polarization" and ``hopping polarization", respectively.
 
\subsection{Lattice polarization}
 
First, for a given ion, the matrix elements are given by
\begin{equation}
  \mel{\vb R, \alpha}{\vb{r}}{\vb R, \beta} = \vb R \delta_{\alpha \beta} + \mel{\vb 0, \alpha}{\vb{r}}{\vb 0, \beta} = \vb R \delta_{\alpha \beta}.
\end{equation}
Indeed, the product of two $d$ or two $p$ orbitals is even under inversion symmetry. This defines the lattice polarization, which reads $\sum_{i} \vb r_i n_{i}$, where $n_{i}= \sum_{\alpha} n_{i, \alpha} =\sum_{\alpha} c\dag_{i, \alpha} c_{i, \alpha}$ for the ion at position $\vb r_i$ (both ligands and TM ions). Moreover, the Mott insulator is overall neutral and has a fixed number of electrons on each equivalent site in the ground state manifold: $n_{\rm TM}$ ($n_{\rm L}$) for TM ions (ligands). The lattice polarization is
\begin{equation}
\label{eq:pol_lat}
  \vb P_{\rm lat} = \sum_{i \in {\rm TM}} \vb r^{\rm TM}_i \delta n_{i} + \sum_{p \in {\rm L}} \vb r^{\rm L}_p \delta n_{p}
\end{equation}
where $\delta n_{i} = n_{i}- n_{{\rm TM} ({\rm L})}$ for TM ions (ligands). This contribution is intrinsic to the tight-binding model. It is directly related to the ``Hubbard current" that we obtain at first order in the vector potential from the Peierls substitution in the hopping integrals.
In the Mott insulating phase, it arises from virtual hopping between neighboring atoms when $t/U$ is finite, so that the ground states are not exact magnetic states but contain a small mixture of polar states.
 After integrating out the $p$-orbital intermediate states, we are left with two types of contributions at the lowest order in $t/U$: the contribution from the TM ions and from the ligands [first and second term in Eq.~\eqref{eq:pol_lat}]. For a TM-TM pair $\langle ij \rangle$, the contribution from the TM ions is
\begin{align}
\label{eq:pollat_pert}
  \vb P_{\textrm{lat}, \langle ij \rangle, \textrm{eff} }^{\textrm{(TM)}} = \mathbb{P} &\Big[\ham^{ij}_{\rm hop} \frac{ \mathbb{Q}_j}{(E_0 - \ham_0)^2}  \ham^{ji}_{\rm hop} \nonumber \\
  & - \ham^{ji}_{\rm hop} \frac{ \mathbb{Q}_i}{(E_0 - \ham_0)^2}  \ham^{ij}_{\rm hop} \Big] \mathbb{P}\, \vb a_{ij},
\end{align}
where $\mathbb{P}$ and $\mathbb{Q}_k$ are the projection operators on the ground state manifold (with energy $E_0$) and polar states with double occupancy at site $k$, respectively. $\ham_0$ is the local Hamiltonian (without hopping), $\ham^{ij}_{\rm hop}$ is the effective hopping Hamiltonian between sites $i$ and $j$, and $\vb a_{ij} = \vb r^{\rm TM}_j - \vb r^{\rm TM}_i$, whose norm is $a$, the lattice spacing. This operator scales as $t_{\textrm{eff}}^2/U^2$ or $t_{pd}^4/(\Delta_{pd}^2 U^2)$. Here $t_{pd}$ generally denotes the TM-L hopping integrals.

The contribution from the ligands comes when we apply $\vb P$ on a polar state with a hole on a ligand. After integrating out the intermediate states, the operator takes the form of an effective hopping,
\begin{align}
\label{eq:lattice_pol_ligand}
	\vb P_{\textrm{lat}, \langle ij \rangle }^{\textrm{(L)}} = \sum_{p} \sum_{\gamma=x,y,z}\frac{t_{ip}^{\alpha \gamma} t_{jp}^{\beta \gamma}} {\Delta_{pd}^2} \vb r^{\rm L}_p \qty( c\dag_{i, \alpha} c_{j, \beta} + c\dag_{j, \beta} c_{i, \alpha} ),
\end{align}
where $p$ labels the ligands between the two TM ions at positions $\vb r^{\rm L}_p$ and $\gamma$ labels their orbitals. This operators scales as $t_{pd}^2/\Delta_{pd}^2$ and its projection onto the ground state manifold is
\begin{align}
\label{eq:pollat_ligand_eff}
\vb P_{\textrm{lat}, \langle ij \rangle , \textrm{eff}}^{\textrm{(L)}}  = \mathbb{P} &\Big[ \vb P_{\textrm{lat},  ij  }^{\textrm{(L)}} \frac{ \mathbb{Q}_j}{E_0 - \ham_0}  \ham^{ji}_{\rm hop} \\
  & + \vb P_{\textrm{lat},  ji  }^{\textrm{(L)}} \frac{ \mathbb{Q}_i}{E_0 - \ham_0}  \ham^{ij}_{\rm hop} \Big] \mathbb{P} +\textrm{H.c.}, \nonumber
 \end{align}
 which, unlike the contributions from TM ions, scales as $t_{pd}^2t_{\textrm{eff}}/(\Delta_{pd}^2 U)$ or $t_{pd}^4/(\Delta_{pd}^3 U)$ and is thus smaller by a factor of $U/\Delta_{pd}$.

\subsection{Hopping polarization}

The remaining matrix elements are
\begin{equation}
  \mel{\vb R, \alpha}{\vb{r}}{\vb R', \beta} = \mel{\vb R, \alpha}{\vb{r}-\vb{r}_0}{\vb R', \beta} = \mel{-\frac{\vb d}{2} , \alpha}{\vb{r}}{\frac{\vb d}{2}, \beta} 
\end{equation}
where $\vb R \ne \vb R'$, $\vb r_0 = (\vb R + \vb R')/2$ and $\vb d = \vb R' - \vb R$ is the vector separating the two ions: either TM-TM (in which case $\norm{\vb d} = a$) or TM-L.
% The hopping polarization is thus obtained from $\mel{\vb r_i, \alpha}{ \vb r }{\vb r_j, \beta}$ integrals
%\begin{equation}
%\label{eq:pol_hop}
%  \vb P_{\rm hop} = \sum_{\langle i,j \rangle} \sum_{\alpha,\beta} \mel{\vb r_i, \alpha}{ \vb r }{\vb r_j, \beta} \qty( c\dag_{i, \alpha} c_{j, \beta} + c\dag_{j, \beta} c_{i, \alpha} ),
%\end{equation}
%where $\vb r_i$ and $\vb r_j$ are the positions of ions $i$ and $j$, respectively (here they refer to both TM ions and ligands). Finally, 
 Anticipating the integrating out of the $p$-orbital states, we only consider the $d$-$d$ and the $p$-$d$ polarization integrals. The behavior under inversion symmetry implies that $\mel{\vb R, \alpha}{ \vb r }{\vb R', \gamma} = \mel{\vb R, \gamma}{ \vb r }{\vb R', \alpha}$ for a $p$-$d$ integral and that $\mel{\vb R, \alpha}{ \vb r }{\vb R', \beta} = -\mel{\vb R, \beta}{ \vb r }{\vb R', \alpha}$ for a $d$-$d$ integral.

After integrating out the intermediate states, we have
\begin{equation}
\label{eq:pol_hop}
  \vb P_{\rm hop} = \sum_{\langle i,j \rangle} \sum_{\alpha,\beta} \vb p_{ij}(\alpha,\beta) \qty( c\dag_{i, \alpha} c_{j, \beta} + c\dag_{j, \beta} c_{i, \alpha} ),
\end{equation}
where $\vb p_{ij}(\alpha,\beta)$ is the effective polarization integral between the TM ions at site $i$ and $j$ with orbitals $\alpha$ and $\beta$, respectively. The projection onto the ground state manifold is obtained similarly to Eq.~\eqref{eq:pollat_ligand_eff}. 

 There are two contributions to $\vb p_{ij}(\alpha,\beta)$: one from direct $d$-$d$ polarization integrals, and a TM-L-TM superexchange-like contribution from $p$-$d$ integrals, so that 
\begin{align}
\label{eq:phopp_eff}
  &\vb p_{ij}(\alpha, \beta) = \mel{\vb r^{\rm TM}_i, \alpha}{ \vb r }{\vb r^{\rm TM}_j, \beta} \\
   &+ \sum_{p} \sum_{\gamma=x,y,z}\frac{
  \mel{\vb r^{\rm TM}_i, \alpha}{ \vb r }{\vb r^{\rm L}_p, \gamma} t_{jp}^{\beta \gamma} + t_{ip}^{\alpha \gamma} \mel{\vb r^{\rm L}_p, \gamma}{ \vb r }{\vb r^{\rm TM}_j, \beta}} 
{\Delta_{pd}} , \nonumber 
\end{align}
where $p$ labels the ligands between the two TM ions and $\gamma$ their orbitals.

Such integrals between $d$ and $p$ orbitals were used in the original KNB derivation \cite{katsura2005spin}. However, in the KNB formalism the integrals are only considered at zeroth order in the interatomic distance (i.e., with $\vb r^{\rm TM}_i = \vb r^{\rm TM}_j$). The outcome of this approximation is that, the finite integrals involving $t_{2g}$ orbitals, $\mel{p_{z}}{y}{d_{yz}}$ and its six permutations, are all equal to each other.

In the following, we derive a more complete scheme to calculate the different integrals for any vector $\vb d$ separating two ions, in a manner that is reminiscent of the Slater-Koster hopping integrals \cite{slater1954simplified}. We find five different symmetry-allowed integrals for a $p$-$d$ pair, and only two for a $d$-$d$ pair.

To summarize, we have two different mechanisms. The lattice polarization has a contribution from the TM ions, which scales as $a \cdot  t^4_{pd}/(\Delta_{pd}^2 U^2)$, and a contribution from the ligands, which scales as $a \cdot t^4_{pd}/(\Delta_{pd}^3 U)$. In the magnetic ground state manifold, the hopping polarization is obtained after one additional virtual hopping and thus scale as $p_{\rm eff} t_{\rm eff}/U$ [where $p_{\rm eff}$ generally represents the effective hopping polarization between two TM sites defined by Eq.~\eqref{eq:phopp_eff}]. In terms of $p$-$d$ integrals, the scaling is $p_{pd}t_{pd}^3/(\Delta_{pd}^2 U)$. Therefore, in the Mott insulating limit where $\Delta_{pd} \gg U$, the dominant contribution is determined by comparing $p_{\rm eff}$ with $a \cdot t_{\rm eff}/U$, or by directly comparing $p_{pd}$ with $a \cdot t_{pd}/U$ if the $p$-$d$ integrals are dominant.

\subsection{Polarization integrals}
The integrals in Eq.~\eqref{eq:pol_hop} are two-center integrals with an extra $l=1$ spherical harmonic at the middle point. A $p$ orbital can be expressed as a linear combination of $p\sigma$ and $p\pi_{\pm}$ functions with respect to the axis $\vb d$ along the bond, and a $d$ orbital can be expressed as a combination of $d\sigma$, $d\pi_{\pm}$, and $d\delta_{\pm}$ functions. Here $\sigma$, $\pi_{\pm}$, and $\delta_{\pm}$ refer to the component of angular momentum along the axis. The different labels directly correspond to the cubic harmonics as shown in Table~\ref{tab:orbitals}.
%Define the five (2+3) d-p symmetry channels and the two d-d (0+2) symmetry channels
\begin{table}[h]
\vspace{5pt}
\begin{tabular}{| c | c | c " c | c | c | c | c |}
  \hline
  $x$ & $y$ & $z$ & $yz$ & $xz$ & $xy$ & $\frac{x^2 - y^2}{2}$ & $\frac{3z^2 - r^2}{2\sqrt{3}}$\\
  \hline
  $\pi_+$ & $\pi_-$ & $\sigma$ & $\pi_-$ & $\pi_+$ & $\delta_+$ & $\delta_-$ & $\sigma$ \\
  \hline
%  $\frac{1}{\sqrt{2}}\qty(Y^{-1}_{1} - Y^1_1)$ & $\frac{i}{\sqrt{2}}\qty(Y^{-1}_{1} + Y^1_1)$ & $Y_1^0$\\
%  \hline  
\end{tabular}
%\quad
%\begin{tabular}{| c | c | c | c | c |}
%  \hline			
%  $yz$ & $xz$ & $xy$ & $\frac{x^2 - y^2}{2}$ & $\frac{3z^2 - r^2}{2\sqrt{3}}$\\
%\hline
%  $\pi_-$ & $\pi_+$ & $\delta_+$ & $\delta_-$ & $\sigma$ \\
%      \hline
%  $\frac{i}{\sqrt{2}}\qty(Y^{-1}_{2} + Y^1_2)$ & $\frac{1}{\sqrt{2}}\qty(Y^{-1}_{2} - Y^1_2)$ & $\frac{i}{\sqrt{2}}\qty(Y^{-2}_{2} - Y^2_2)$& $\frac{1}{\sqrt{2}}\qty(Y^{-2}_{2} + Y^1_2)$  & $Y_2^0$  \\
%  \hline  
%\end{tabular}
\caption{Cubic harmonics of $p$ and $d$ orbitals expressed with respect to their component of angular momentum along the $z$ axis.}
\vspace{-10pt}
\label{tab:orbitals}
\end{table}
For $p$-$d$ integrals, five symmetry channels are allowed. They can be separated into longitudinal ($\vb P$ along the bond) and transverse ($\vb P$ orthogonal to the bond) components.
%, referring to the direction of $\vb r$ considered. 
To lighten the notation, the center of the atomic orbitals is omitted and assumed to be $-\vb d/2$ ($\vb d/2$) for the bra (ket) state. The two longitudinal integrals are
\begin{align}
  P^{\parallel}_{pd\sigma}&=\mel{p_{\sigma}}{ \sigma }{d_{\sigma}} \\
  P^{\parallel}_{pd\pi}&=\mel{ p_{\pi_{\pm}}}{ \sigma }{  d_{\pi_{\pm}}}
\end{align}
and the three transverse integrals are
\begin{align}
  P^{\perp}_{p\pi d\sigma}&=\mel{ p_{\pi_{\pm}}}{ \pi_{\pm} }{  d_{\sigma}} \\
  P^{\perp}_{p\sigma d\pi}&=\mel{ p_{\sigma}}{ \pi_{\pm} }{  d_{\pi_{\pm}}} \\
  P^{\perp}_{p\pi d\delta}&=\mel{ p_{\pi_{\pm}}}{ \pi_{\mp} }{  d_{\delta_{+}}} \nonumber \\
  &=\pm \mel{ p_{\pi_{\pm}}}{ \pi_{\pm} }{  d_{\delta_{-}}}
\end{align}
Here $\sigma$ and $\pi_{\pm}$ denote the projections of $\vb r$ on the $\sigma$ and $\pi_{\pm}$ axes.

For the $d$-$d$ integrals, there are only two transverse components,
\begin{align}
  P^{\perp}_{d\sigma d\pi}
  &=\mel{ d_{\sigma}}{ \pi_{\pm} }{  d_{\pi_{\pm}}}\nonumber \\
  &= -\mel{ d_{\pi_{\pm}}}{ \pi_{\pm} }{  d_{\sigma}} \\
  P^{\perp}_{d\pi d\delta}
   &=\mel{ d_{\pi_{\pm}}}{ \pi_{\mp} }{  d_{\delta_+}}\nonumber \\
   &=\pm \mel{ d_{\pi_{\pm}}}{ \pi_{\pm} }{  d_{\delta_-}}\nonumber \\
   &=-\mel{ d_{\delta_+}}{ \pi_{\mp} }{  d_{\pi_{\pm}}}\nonumber \\
   &=\mp \mel{ d_{\delta_-}}{ \pi_{\pm} }{  d_{\pi_{\pm}}}.
\end{align}

\begin{figure}[t]
\centering
\begin{minipage}{.245\textwidth}
  \centering
  \includegraphics[width=\linewidth]{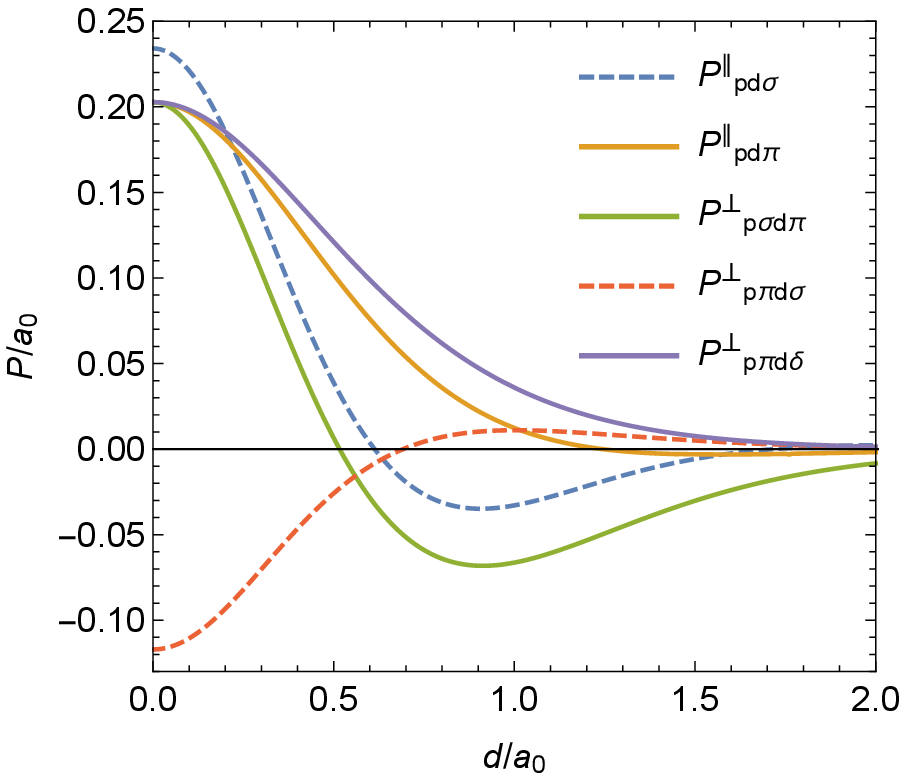}
\end{minipage}%
\begin{minipage}{.245\textwidth}
  \centering
  \includegraphics[width=\linewidth]{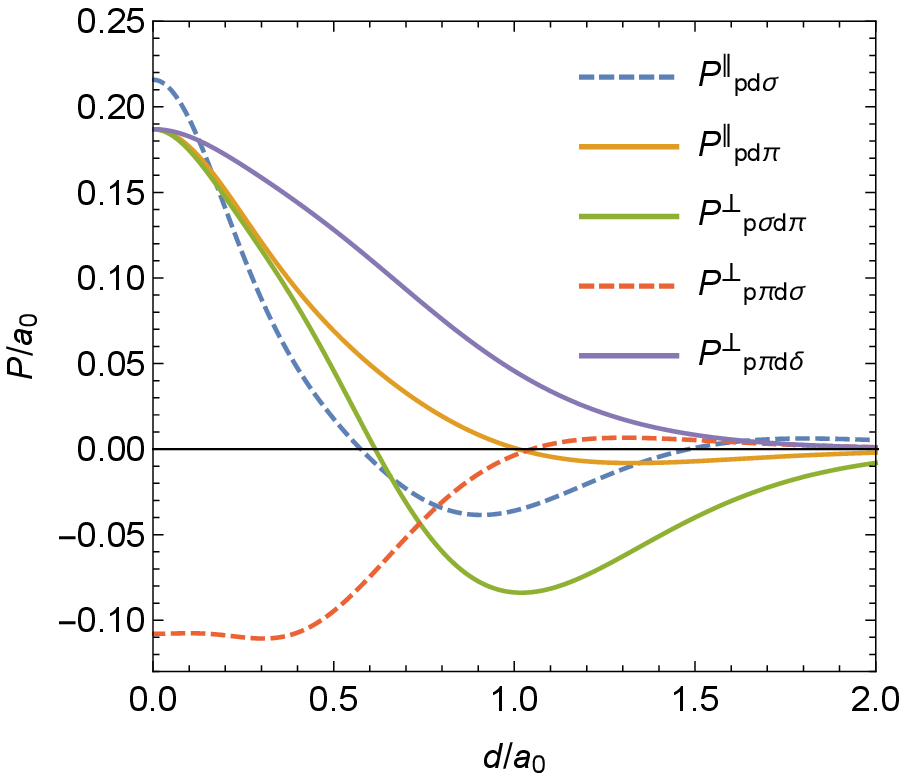}
\end{minipage}

\begin{minipage}{.245\textwidth}
  \centering
  \includegraphics[width=\linewidth]{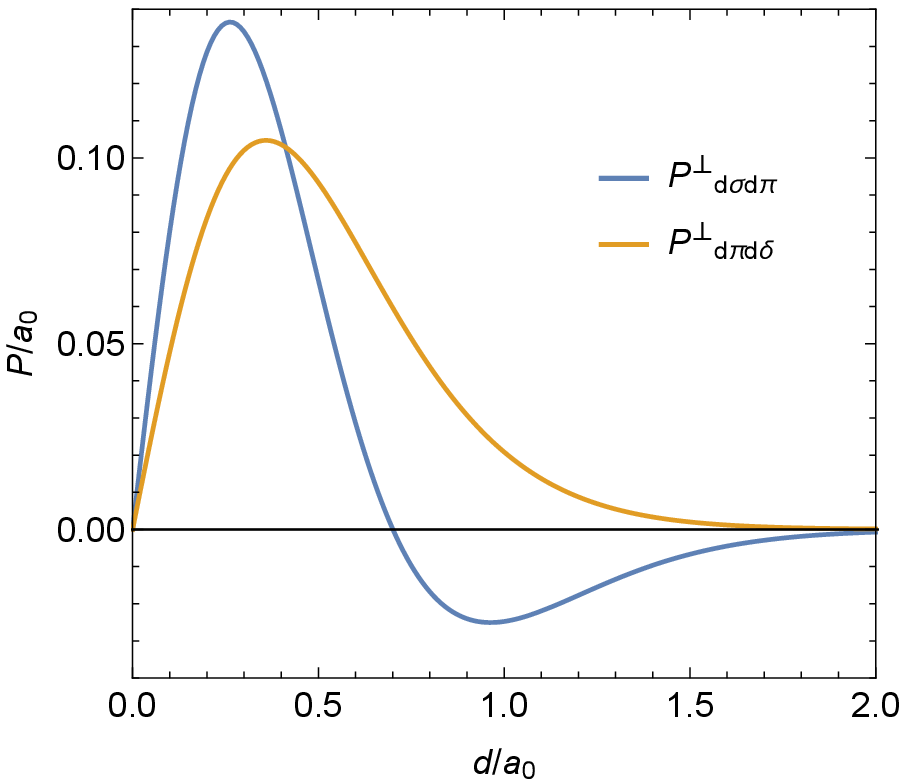}
\end{minipage}%
\begin{minipage}{.245\textwidth}
  \centering
  \includegraphics[width=\linewidth]{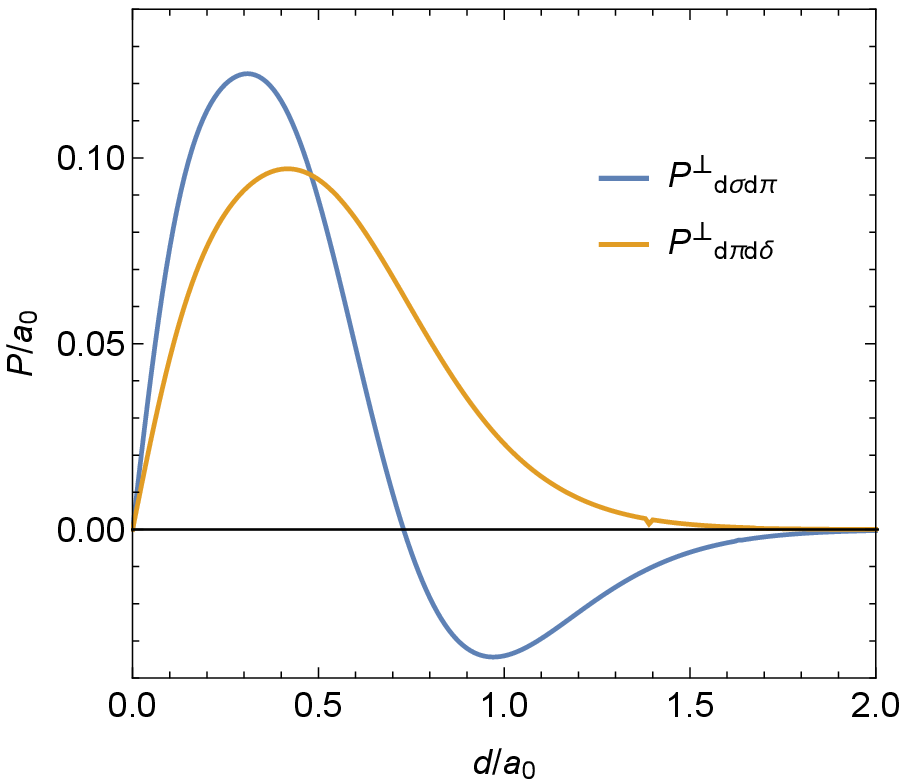}
\end{minipage}
%\caption{The different $p$-$d$ polarization integrals as a function of the inter-atomic distance $d$, in units of $a_0$, the Bohr radius. The dashed lines are the integrals not relevant to the $t_{2g}$ orbitals when $\vb d$ is along one of the main axes. Left: atomic orbitals of chromium and oxygen. Right: atomic orbitals of ruthenium and chloride.}
\caption{The different $p$-$d$ (top) and $d$-$d$ (bottom) polarization integrals as a function of the atomic spacing $d$, in units of $a_0$, the Bohr radius. The cases of chromium and oxygen atoms (left) and ruthenium and chloride atoms (right) are plotted using hydrogenlike atomic orbitals.}
\label{fig:integrals}
\end{figure}

From those expressions, we can then calculate the polarization integrals in a fixed basis for $\vb d$ pointing in an arbitrary direction. Using the notation of Slater and Koster \cite{slater1954simplified}, we denote the cosine angles of $\vb d$ as $(l,m,n)$. For example we have
\begin{align}
  \mel{p_{x}}{x}{d_{xy}} =& \sqrt{3} l^3 m P^{\parallel}_{pd\sigma} + l m \left(1-2l^2\right)P^{\parallel}_{pd\pi}  \nonumber \\
   &+  \sqrt{3} l m
   \left(1-l^2\right) P^{\perp}_{p\pi d\sigma}
   + l m \left(1-2l^2\right) P^{\perp}_{p\sigma d\pi} \nonumber \\
    &-l m
   \left(1-l^2\right)P^{\perp}_{p\pi d\delta}.
\end{align}
Other integrals are found in Table~\ref{tab:pol} in Appendix~\ref{sec:polApendix}.
In particular, for a bond along the $z$ axis, the $t_{2g}$-$p$ integrals of the KNB theory split into three different possible values:
\begin{align}
\label{eq:KNBsplit}
  \mel{p_x}{z}{d_{xz}} = \mel{p_y}{z}{d_{yz}} = P^{\parallel}_{pd\pi}\nonumber \\
  \mel{p_z}{y}{d_{yz}} = \mel{p_z}{x}{d_{xz}} = P^{\perp}_{p\sigma d\pi} \nonumber \\
  \mel{p_x}{y}{d_{xy}} = \mel{p_y}{x}{d_{xy}} = P^{\perp}_{p\pi d\delta}
\end{align}
%
%
% depending on the relative orientation of the bond (assuming it is along one of the $\sigma$ or $\pi_{\pm}$ axis):  $P^{\parallel}_{pd\pi}$, $P^{\perp}_{p\sigma d\pi}$ and $P^{\perp}_{p\pi d\delta}$
In the limit where $d = \norm{\vb d} \rightarrow 0$, the three integrals of Eq.~\eqref{eq:KNBsplit} are identical and can be roughly calculated using the orbitals of the hydrogen like atom \cite{katsura2005spin}. Moreover, the $d$-$d$ integrals vanish due to their behavior under inversion symmetry.
The integrals have been numerically calculated as a function of the interatomic distance $d$ using hydrogenlike atomic orbitals. The results are shown in Fig.~\ref{fig:integrals}. The hydrogenlike atomic orbitals of chromium ($3d$ orbitals with $Z=24$) and oxygen ($2p$ orbitals with $Z=8$), and of ruthenium ($4d$ orbitals with $Z=44$) and chloride ($3p$ orbitals with $Z=17$) have been used (left and right figures, respectively). For a typical interatomic distance of $2 \text{\AA} \sim 3.8 a_0$ ($a_0$ is the Bohr radius), the $p$-$d$ integrals are considerably smaller than the zeroth order approximation used in the KNB theory by a factor of $\sim 10^{4-6}$. 

It is important to note that in general, the hydrogenlike atom model is not a good approximation. In particular, when the system is close to the Mott transition, or close to a molecular-insulating regime, the wave functions might be spread over several sites with significant overlap. This is especially true for the $4d$ or $5d$ compounds considered in this paper. 
 In such a case, the polarization integrals should be larger (as are the tight-binding hopping integrals) and first-principle methods should be used to evaluate the atomic orbitals. Nevertheless, the symmetry considerations (the classification into different symmetry channels) still apply.

%\begin{figure}
%\centering
%	\includegraphics[width=0.40\textwidth]{./integrals_numerics.eps}
%	\caption{The different $p$-$d$ polarization integrals as a function of the interatomic distance $d$, in units of $a_0$, the Bohr radius. The dashed lines are the integrals not relevant to the $t_{2g}$ orbitals when $\vb d$ is along one of the main axes.}
%\label{fig:pdintegrals}
%\end{figure}

%\begin{figure}
%\centering
%	\includegraphics[width=0.40\textwidth]{./ddintegrals_numerics.eps}
%	\caption{The different $d$-$d$ polarization integrals as a function of the inter-atomic distance $d$, in units of $a_0$, the Bohr radius.}
%\label{fig:ddintegrals}
%\end{figure}

\section{Microscopic mechanisms}
\label{sec:mechanism}

We are now looking explicitly at the mechanisms behind the allowed couplings in terms of the microscopic lattice and hopping polarization defined in Sec.~\ref{sec:polarization}. For each allowed coupling constants $m_{1-5}$ in Eqs.~\eqref{eq:pol_allowed_noCF}, \eqref{eq:pol_allowed_CF}, \eqref{eq:polcorner_allowed_noCF}, and \eqref{eq:polcorner_allowed_CF}, we separate the contributions
%\begin{align}
% 	\vb P_{\textrm{eff}} = &  m_1\qty[ \vu u \cdot \left( \S_i \times \S_j \right) ]\vu u \nonumber \\
% 	 + & \qty [m_2  \left( \S_i \times \S_j \right) \cdot \vu v  + m_4  \left( \S_i \times \S_j \right) \cdot \vu w ] \vu v \nonumber \\
% 	 + & \qty[ m_5  \left( \S_i \times \S_j \right) \cdot \vu v  + m_3 \left( \S_i \times \S_j \right) \cdot \vu w ] \vu w,
%\end{align}
as
\begin{equation}
\label{eq:decomp}
  m_i = a \mathbb{A}_i + \mathbb{B}_i,
\end{equation}
where $\mathbb{A}_i$ and $\mathbb{B}_i$ are the coupling constants resulting from the $\vb P_{\textrm{lat}}$ and $\vb P_{\textrm{hop}}$, respectively. Here $a$ is the lattice spacing between two TM ions so that the contributions $\mathbb{A}_i$ are unitless, while the contributions $\mathbb{B}_i$ have units of distance coming from the polarization integrals. The results are summarized in Table~\ref{tab:results_edge} and shown in Figs.~\ref{fig:numericsA} and \ref{fig:numericsB} for the edge-sharing geometry, and are summarized in Table~\ref{tab:results_corner} for the corner-sharing geometry.

\subsection{Edge-sharing geometry}
\label{sec:face_mech}
In the edge-sharing geometry, the five coupling constants $m_{1-5}$ in Eq.~\eqref{eq:pol_allowed_CF} are defined through
\begin{align}
\label{eq:coupling_cst_edge}
 	\vb P_{\textrm{eff}} = &  m_1\qty[ \vu u \cdot \left( \S_i \times \S_j \right) ]\vu u \nonumber \\
 	 + & \qty [m_2  \left( \S_i \times \S_j \right) \cdot \vu v  + m_4  \left( \S_i \times \S_j \right) \cdot \vu w ] \vu v \nonumber \\
 	 + & \qty[ m_5  \left( \S_i \times \S_j \right) \cdot \vu v  + m_3 \left( \S_i \times \S_j \right) \cdot \vu w ] \vu w.
\end{align} 

\subsubsection{Lattice polarization}

In order to calculate the contribution from $\vb P_{\textrm{lat}}$ in Eq.~\eqref{eq:pol_lat}, we consider the subsystem made of one bond depicted in Fig.~\ref{fig:geometries}(b): two TM ions at sites $i$ and $j$ separated by $a \vu u$, and two ligands also separated by a distance $a$ when $\phi = 90^{\circ}$. 
All the calculations done in this section are extended to the $\phi \neq 90^{\circ}$ case in Appendix \ref{sec:edge_angle}. In the subsystem, $\sum_k \delta n_k =0$ and we can choose the origin at the center of the bond. Then, the contribution from the TM ions is always in the $\vu u$ direction, and the contribution from the ligands is perpendicular to it. Without trigonal distortion, the latter is exactly in the $\vu v$ direction.

Without trigonal distortion, we see in Eq.~\eqref{eq:pol_allowed_noCF} that from symmetry considerations, $m_1=0$. Hence, the contribution to $\vb P_{\rm lat}$ along the bond coming from the TM ions vanishes due to the high $D_{2h}$ symmetry of the bond.
More intuitively, this can be understood from the original microscopic Hamiltonian \eqref{eq:ham_general}. With the trigonal CF $\Delta =0$, the on-site Hamiltonian is $O(3)$ rotationally invariant because we neglect the $e_g$ orbitals. The bond symmetry is encoded in the hopping matrix, which has three eigenvectors corresponding to the three $C_2$ axes of the $D_{2h}$ group: $\vu u$, $\vu v$, and $\vu w$. 
The on-site energy eigenstates are grouped into three Kramers pairs (or three states with a pseudospin variable). Because SOC is rotationally invariant, the hopping is completely diagonal with respect to the pseudospins. As the polarization only connects states with different parities, the contribution from $\vb P_{\rm lat}^{\rm (TM)}$ along the bond must vanish.
	When $\Delta \ne 0$, the hopping is no longer diagonal with respect to the pseudospins and a finite contribution is possible.

Additionally, in the specific case where $(\vu x + \vu y + \vu z)/\sqrt{3}$ is an eigenvector of the hopping matrix (i.e., the CF Hamiltonian and the hopping matrix are simultaneously diagonalizable), the system recovers an accidental $D_{2h}$ symmetry and the contribution vanishes. This happens exactly when $t_1 + t_2 - t_3 - t_4 = 0$. 
%Note that this is only true because we neglected the $e_{g}$ orbitals.

The contribution to $m_1$ in Eq.~\eqref{eq:pol_allowed_CF} can be found numerically with exact diagonalization of the two-site system or with perturbation theory. Using perturbation theory at second order in $\ham_{\rm hop}$, as explained in Sec.~\ref{sec:polarization}, we find a contribution to $m_1$,
\begin{equation}
\label{eq:poleff_metal}
 	\vb P_{\textrm{lat, eff}}^{\textrm{(TM)}}= \mathbb{A}_1\qty[ \vu u \cdot \left( \S_i \times \S_j \right) ] a \vu u.
\end{equation}
From Eq.~\eqref{eq:pollat_pert},
\begin{equation}
  \mathbb{A}_1\qty[ \vu u \cdot \left( \S_i \times \S_j \right) ] =  P_{ij} - P_{ji},
\end{equation}
where
\begin{equation}
\label{eq:pol2nd}
  P_{ij} = \mathbb{P} \ham^{ij}_{\rm hop} \frac{ \mathbb{Q}_j}{(E_0 - \ham_0)^2}  \ham^{ji}_{\rm hop} \mathbb{P}.
\end{equation}
%and Eq.~\eqref{eq:pollat_pert}, perturbation theory at second order in the effective TM-TM hopping is expressed as
%\begin{equation}
%  \vb P_{\textrm{lat}}^{\textrm{(TM)}} = \left( P_{ij} - P_{ji} \right) \frac{a}{2} \vu u,
%\end{equation}
%with
%\begin{equation}
%\label{eq:pol2nd}
%  P_{ij} = \mathbb{P} \ham^{ij}_{\rm hop} \frac{ \mathbb{Q}}{(E_0 - \ham_0)^2}  \ham^{ji}_{\rm hop} \mathbb{P},
%\end{equation}
Here, $\ham_0 = \ham_{\textrm{SOC}} + \ham_{\textrm{CF}} + \ham_{\textrm{int}}$.
Including the trigonal distortion, the calculation is too cumbersome to be performed analytically, but we can treat $\ham_{\textrm{CF}}$ as an additional perturbation. At the lowest order (second order in the hopping and first order in $\Delta$),
\begin{align}
\label{eq:pol3rd}
  P_{ij} =& \Big[-\frac{2}{3\lambda} \mathbb{P}_{\frac{1}{2}} \ham^{ij}_{\rm hop} \frac{\mathbb{Q}_j}{(E_0 - \ham'_0)^2} \ham^{ji}_{\rm hop}  \mathbb{P}_{\frac{3}{2}}  \ham_{\rm CF} \mathbb{P}_{\frac{1}{2}}\nonumber \\
   & +\mathbb{P}_{\frac{1}{2}} \ham^{ij}_{\rm hop} \frac{\mathbb{Q}_j}{(E_0 - \ham'_0)^2} \ham_{\rm CF} \frac{\mathbb{Q}_j}{E_0 - \ham'_0} \ham^{ji}_{\rm hop}   \mathbb{P}_{\frac{1}{2}} \Big] \nonumber \\
    &+ \rm{H.c.},
\end{align}
where $\mathbb{P}_{\frac{1}{2}}$ and $\mathbb{P}_{\frac{3}{2}}$ are the projection operators on the $J=1/2$ states and the $J = 3/2$ states, respectively. In Eq.~\eqref{eq:pol3rd}, $\ham'_0 = \ham_{\textrm{SOC}} + \ham_{\textrm{int}}$.

\begin{figure*}
\centering
%\hspace*{-0.7cm}
\includegraphics[width=0.9\textwidth]{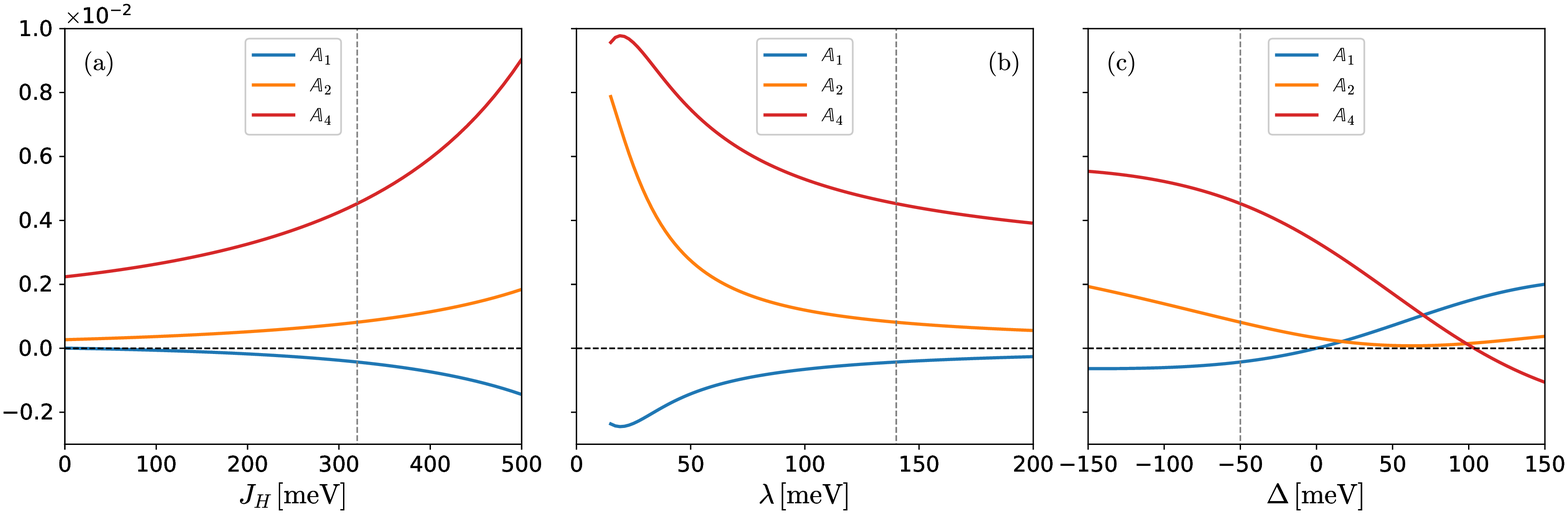}
	\caption{Numerical evaluation of $\mathbb{A}_1$, $\mathbb{A}_2$, and $\mathbb{A}_4$ in Eqs.~\eqref{eq:decomp} and \eqref{eq:pollat_ligand} using exact diagonalization as a function of (a) $J_H$, (b) $\lambda$, and (c) $\Delta$. The nonvarying physical parameters are chosen as $U=2310$ meV, $J_H=320$ meV, $\lambda = 140$ meV, $\Delta=-50$ meV, $t_1=50$ meV, $t_2 = 160$ meV, $t_3 = -150$ meV, and $t_4 = -20$ meV, typical for $\alpha$-RuCl$_3$.}
\label{fig:numericsA}
\end{figure*}

The full expression of $\mathbb{A}_1$ is large and given in Appendix~\ref{sec:expression_edge}. It can be written as
\begin{align}
\label{eq:Aconstgeneral}
  \mathbb{A}_1 = %\sqrt{2} \frac{64}{81}
   \Delta (t_1+t_2-t_3-t_4) \frac{J_H}{\lambda}
 \times \frac{P}{Q},
 \end{align}
where the $P$ and $Q$ are polynomials given in Appendix~\ref{sec:expression_edge}. In the $U \gg J_H,\lambda$ limit, we find
\begin{equation}
	  \lim_{\substack{J_H/U  \rightarrow 0 \\ \lambda/U \rightarrow 0}} \mathbb{A}_1 = \sqrt{2} \frac{64}{81}\Delta (t_1+t_2-t_3-t_4)(t_1+t_2+t_3)  \frac{J_H}{\lambda U^3}.
	  \label{eq:Aconstgeneral2}
\end{equation}
Interestingly, the unitless constant $\mathbb{A}_1$ vanishes when $J_H=0$. This can be seen in Fig.~\ref{fig:numericsA}(a) as well.
The ME coupling in Eq.~\eqref{eq:poleff_metal} was used in Ref.~\cite{bolens2018mechanism} to explain the subgap optical conductivity of Kitaev materials.
%For clarity, we only show $\mathbb{A}_1$ at first order in $J_H$,
%\begin{align}
%  \label{eq:Aconstgeneral}
%  \mathbb{A}_1 =&  \sqrt{2} \Delta (t_1+t_2-t_3-t_4)\frac{128}{81}  \frac{1}{ \lambda  U^2 (3 \lambda +2
%   U)^4} \times \nonumber \\
%  & J_H \Big[ 8 U^3
%   (t_1+t_2+t_3)+3 \lambda  U^2 (11 t_1+7 t_2+9 t_3) \nonumber \\
%   &  +24 \lambda ^2 U (2 t_1+t_3) +9 \lambda ^3 (2t_1+t_3) \Big]  +O(J_H^2).
%\end{align}

\begin{figure*}
\centering
%\hspace*{-0.7cm}
\includegraphics[width=0.9\textwidth]{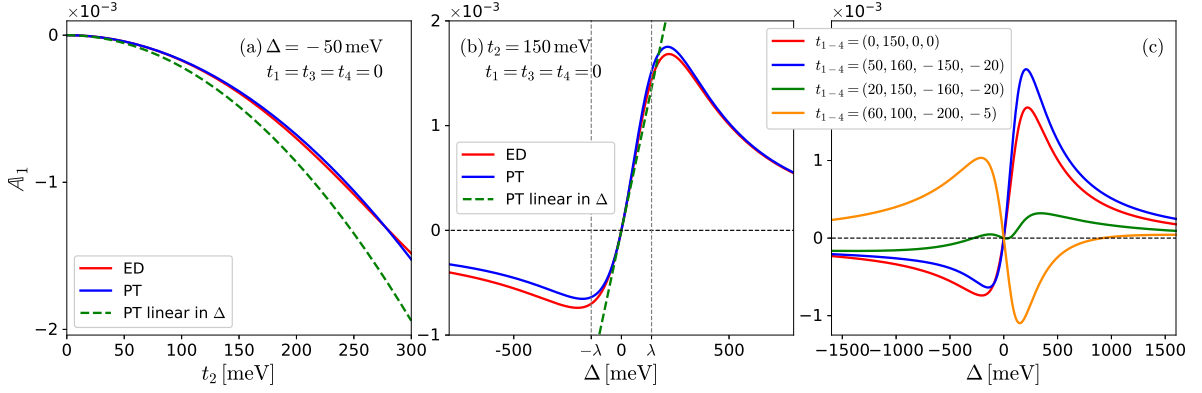}
	\caption{Numerical evaluation of $\mathbb{A}_1$ in Eq.~\eqref{eq:poleff_metal}. Results using exact diagonalization (ED) (red line), perturbation theory (PT) exact in $\Delta$ (blue line), and perturbation theory linear in $\Delta$ (green dashed line) are plotted in (a) and (b), for typical values of the physical parameters: $U=2310$ meV, $J_H=320$ meV, $\lambda = 140$ meV, and we set $t_1=t_3=t_4=0$. It is plotted (a) as a function of $t_2$ with $\Delta = -50$ meV and (b) as a function of $\Delta$ with $t_2=150$ meV. (c) shows ED calculation of $\mathbb{A}_1$ as a function of $\Delta$ for different sets of values $t_{1-4}$ taken from the literature \cite{winter2016challenges, wang2017theoretical}. The values are indicated in meV units.}
\label{fig:ED_PT}
\end{figure*}

The calculation of $\mathbb{A}_1$ was also performed using perturbation exact in $\Delta$ and using exact diagonalization on a two-site system. 
In Fig.~\ref{fig:ED_PT}, we plot $\mathbb{A}_1$ calculated with the three methods: exact diagonalization, perturbation theory (quadratic in the hopping) and from Eq.~\eqref{eq:Aconstgeneral} (quadratic in the hopping and linear in the CF distortion). The physical parameters were set to $U=2310$ meV, $J_H=320$ meV, $\lambda = 140$ meV, which correspond to typical values for $\alpha$-RuCl$_3$ \cite{winter2016challenges, wang2017theoretical}. In Figs.~\ref{fig:ED_PT}(a) and \ref{fig:ED_PT}(b), we set $t_1=t_3=0$ and calculate $\mathbb{A}_1$ as a function of $t_2$ and $\Delta$, respectively. For $\alpha$-RuCl$_3$, the trigonal CF distortion $\Delta$ is typically somewhere between $-15$ meV and $-70$ meV (not so small compared to $\lambda$). In Fig.~\ref{fig:ED_PT}(c), we also consider finite values for $t_1$, $t_3$, and $t_4$, and plot exact diagonalization results for different sets of realistic values. We find that for typical values of the hopping amplitudes, the perturbation theory calculations very well reproduce the exact diagonalization calculations. Moreover, the perturbation theory linear in $\Delta$ is, in most cases, a surprisingly good approximation up to relatively large values of $\Delta$ as long as $\abs{\Delta} < \lambda$ (which is usually the case in $4d$ and $5d$ materials due to strong SOC). Interestingly, the behavior depends considerably on the hopping integrals and their relative amplitudes. In particular, the sign of $\mathbb{A}_1$ changes as $\abs{t_3/t_2}$ increases. Finally, we properly observe in Figs.~\ref{fig:ED_PT}(b) and \ref{fig:ED_PT}(c) that both SOC and the trigonal CF are required in order have a finite $\vb P$ along the bond. Indeed, $\mathbb{A}_1 \rightarrow 0$ for $\Delta \rightarrow \pm \infty$, which is equivalent to $\lambda \rightarrow 0$.

The second contribution to $\vb P_{\textrm{lat}}$, from the ligands, is obtained from Eq.~\eqref{eq:lattice_pol_ligand} and is just the effective hopping matrix elements arising from the TM-L-TM processes, but with opposite sign for the two ligands so that the matrix is antisymmetric. Without trigonal distortion, it is oriented along $\vu v$ so that
\begin{equation}
 \label{eq:pollat_ligand}
\vb P_{\textrm{lat}}^{\textrm{(L)}} = \frac{t_{pd\pi}^2}{\Delta_{pd}^2} \qty[ \vb c\dag_i
  \begin{pmatrix}
  	0 & -1 & 0 & \\
  	1 & 0 & 0 \\
  	0 & 0 & 0 \\
  \end{pmatrix}
  \vb c_j + \textrm{H.c.} ] \frac{a}{2} \vu v.
\end{equation}
%begin{equation}
%\frac{t_{pd\pi}^2}{\Delta_{pd}}\sum_{\alpha, \beta} \begin{pmatrix}
%  	0 & -1 & 0 & \\
%  	1 & 0 & 0 \\
%  	0 & 0 & 0 \\
%  \end{pmatrix}_{\alpha \beta}  ( c\dag_{i,\alpha} c_{j, \beta} + \textrm{H.c.} ) \vu v
%\end{equation}
%(note: i-p :x bond = + (xz=Y), y bond = -(yz=X), j-p: x bond = - (xz=Y), y bond = + (yz=X))
%As before, the hopping integral $t_{pd\pi}$ itself also depends on $\phi$ due to the change of the TM-ligand distance. 
Equation~\eqref{eq:pollat_ligand} has the same form as the hopping polarization terms considered in the next subsection. We already know from symmetry that it eventually yields
\begin{align}
\label{eq:poleff_ligand}
 	\vb P_{\textrm{lat},\textrm{eff}}^{\textrm{(L)}} =& \qty [\mathbb{A}_2  \vu v \cdot \left( \S_i \times \S_j \right) + \mathbb{A}_4  \vu w \cdot \left( \S_i \times \S_j \right) ] a \vu v \nonumber \\
 	 +& \qty[ \mathbb{A}_5  \vu v \cdot \left( \S_i \times \S_j \right)  + \mathbb{A}_3 \vu w \cdot \left( \S_i \times \S_j \right) ] a \vu w,
\end{align}
where only $\mathbb{A}_4 \neq 0$ without trigonal CF. Furthermore, it is safe to neglect the displacement of the ligands along $\vu w$ in which case $\mathbb{A}_5 = \mathbb{A}_3 = 0$. The constants are explicitly calculated in the next subsection.

In Fig.~\ref{fig:numericsA}, we plot the coupling constants $\mathbb{A}_{1}$, $\mathbb{A}_2$, and $\mathbb{A}_4$, calculated with exact diagonalization on a two-site cluster without any approximations, as a function of $J_H$, $\lambda$, and $\Delta$ in Figs.~\ref{fig:numericsA}(a), \ref{fig:numericsA}(b), and \ref{fig:numericsA}(c), respectively. Note that we cannot decrease $\lambda$ below a certain threshold in Fig.~\ref{fig:numericsA}(b) because the pseudospin 1/2 are not well defined for $\Delta <0$ when $\lambda = 0$ (the local ground state is fourfold degenerate).

\begin{figure*}
\centering
%\hspace*{-0.7cm}
\includegraphics[width=0.9\textwidth]{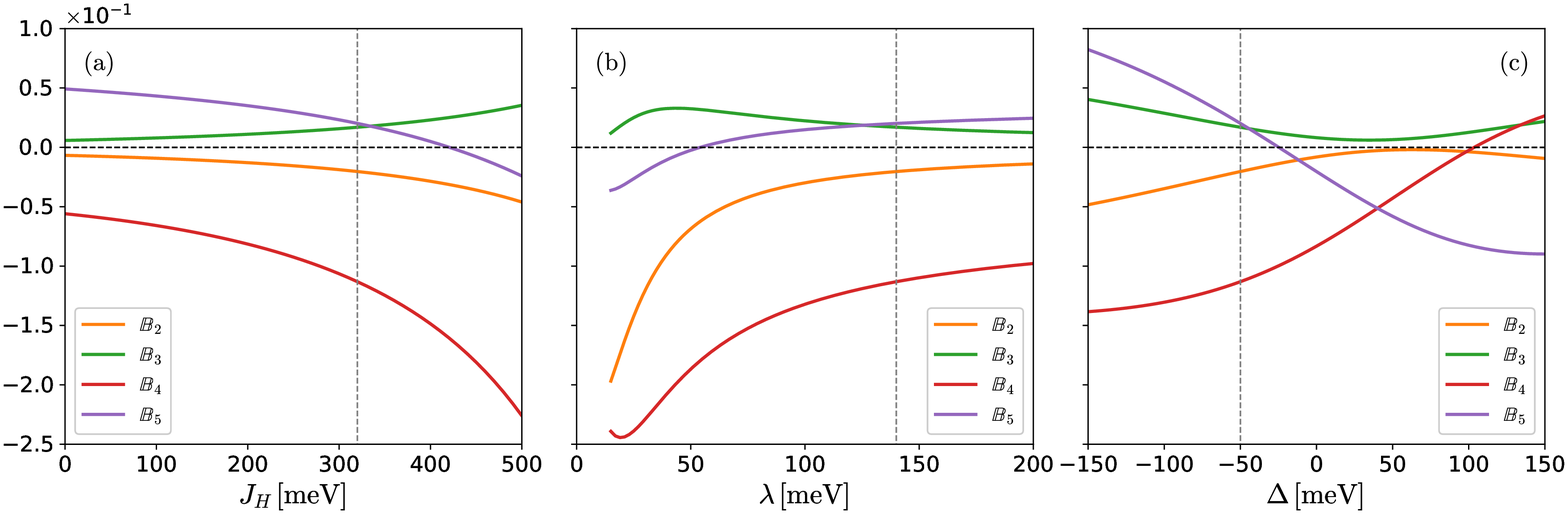}
	\caption{Numerical evaluation of $\mathbb{B}_{2-4}$ in Eq.~\eqref{eq:decomp} using exact diagonalization as a function of (a) $J_H$, (b) $\lambda$, and (c) $\Delta$. The nonvarying physical parameters are chosen as $U=2310$ meV, $J_H=320$ meV, $\lambda = 140$ meV, $\Delta=-50$ meV, $t_1=50$ meV, $t_2 = 160$ meV, $t_3 = -150$ meV, and $t_4 = -20$ meV, typical for $\alpha$-RuCl$_3$. The coupling constants are plotted in units of $b_5$ (see Eq.~\eqref{eq:pol_int_b45}) where we assumed $b_4 = - b_5$ for simplicity.}
\label{fig:numericsB}
\end{figure*}

\clearpage

\begin{table}[t]
\caption{Coupling constants defined in Eqs.~\eqref{eq:pol_allowed_CF} and \eqref{eq:decomp} for the edge-sharing geometry in the $J_H/U \rightarrow 0$ limit (and $\lambda/U \rightarrow 0$ for $\mathbb{A}_1$). The CF is included through $\ham_{\rm CF}$ but the displacement of the ligands is neglected. The coefficients $b_4$ and $b_5$ are given in terms of the polarization integrals in Eq.~\eqref{eq:pol_int_b45}.}
\renewcommand{\arraystretch}{1.5}
\begin{tabular}{| c c l |}
  \hline
  $\mathbb{A}_1$ & $=$ & $\sqrt{2} \frac{64}{81} \Delta (t_1+t_2-t_3-t_4)(t_1+t_2+t_3) \frac{J_H}{\lambda U^3}$ \\
  $\mathbb{A}_2$ & $=$ & $-\frac{t_{pd\pi}^2}{\Delta_{pd}^2} \sqrt{2}\frac{16}{81} \Delta \frac{2t_1 +t_3}{\lambda U}$  \\
  $\mathbb{A}_3$ & $=$ & $0$ \\
  $\mathbb{A}_4$ & $=$ & $\frac{t_{pd\pi}^2}{\Delta_{pd}^2} \qty[\frac{8}{9} \frac{2t_1 + t_3 }{U} + \frac{64}{81} \Delta \frac{t_2 + 2 t_4}{\lambda U}]$ \\
  $\mathbb{A}_5$ & $=$ & $0$ \\
  \hline
  $\mathbb{B}_1$ & $=$ & $0$ \\
  $\mathbb{B}_2$ & $=$ & $b_4 \sqrt{2}\frac{32}{81} \Delta \frac{2t_1 +t_3}{\lambda U}$ \\
  $\mathbb{B}_3$ & $=$ & $b_5 \sqrt{2} \frac{32}{81} \Delta \frac{2t_1 +t_3}{\lambda U} $ \\
  $\mathbb{B}_4$ & $=$ & $-b_4\qty[\frac{16}{9} \frac{2t_1 + t_3 }{U} + \frac{128}{81} \Delta \frac{t_2 + 2 t_4}{\lambda U}]$ \\
  $\mathbb{B}_5$ & $=$ & $b_5 \qty[-\frac{16}{9} \frac{2t_1 + t_3 }{U} + \frac{32}{81} \Delta \frac{2t_1 - 4t_2 + t_3 - 8t_4}{\lambda U}]$ \\
  \hline
  \end{tabular}
\label{tab:results_edge}
\end{table}

\subsubsection{Hopping polarization}

The contribution from the microscopic operator~\eqref{eq:pol_hop} can be decomposed in terms of the hopping operators
\begin{align}
\label{eq:matrices}
  &\hat M_{u} =
   \frac{1}{\sqrt{2}} \vb c_i\dag \begin{pmatrix}
  	0 & 0 & 1 & \\
  	0 & 0 & 1 \\
  	-1 & -1 & 0 \\
  \end{pmatrix} \vb c_j + \textrm{H.c.}, \nonumber \\
    &\hat M_{v} = \frac{1}{\sqrt{2}}
   \vb c_i\dag \begin{pmatrix}
  	0 & 0 & -1 & \\
  	0 & 0 & 1 \\
  	1 & -1 & 0 \\
  \end{pmatrix}\vb c_j + \textrm{H.c., and} \nonumber \\
  &\hat M_{w} =
  \vb c_i\dag \begin{pmatrix}
  	0 & 1 & 0 & \\
  	-1 & 0 & 0 \\
  	0 & 0 & 0 \\
  \end{pmatrix}\vb c_j + \textrm{H.c.},
\end{align}
where $\hat M_n$ acts in the same representations as $\qty( \S_i \times \S_j) \cdot \vu n$ for $\vu n=\vu u, \vu v$, and $\vu w$. Thus, in the $C_{2h}$ bond symmetry, 
\begin{align}
  \vb P_{\textrm{hop}} =& \qty[b_1 \hat M_u] \vu u + \qty[ b_2 \hat M_v + b_4 \hat M_w] \vu v + \qty[ b_5 \hat M_v  + b_3 \hat M_w] \vu w,
\end{align}
and only $b_4$ and $b_5$ survive when the displacement of the ligands away from the $D_{2h}$ symmetry is neglected.
When computed in the ground state manifold of the Hamiltonian without trigonal distortion, each $\hat M_n$ creates a $\qty( \S_i \times \S_j) \cdot \vu n$ term due to the $D_{2h}$ symmetry. With the addition of $\ham_{\textrm{CF}}$, using the $C_{2h}$ character table~\ref{tab:C2}, we infer that the $b_2$ and $b_4$ terms mix and contribute to both $\mathbb{B}_2$ and $\mathbb{B}_4$. Similarly the $b_5$ and $b_3$ terms mix and contribute to both $\mathbb{B}_5$ and $\mathbb{B}_3$. Finally, the $b_1$ term contributes to $\mathbb{B}_1$. At first order in the hopping Hamiltonian [see Eq.~\eqref{eq:pollat_ligand_eff}] we have
\begin{align}
  \hat M_{n, \textrm{eff}} = \mathbb{P} \hat M_n  \frac{1}{E_0 - \ham_0} \ham_{\textrm{hop}} \mathbb{P} +\textrm{H.c.}.
  \label{eq:Mpert}
\end{align}
Without trigonal distortion, Eq.~\eqref{eq:Mpert} results in $\hat M_{n, \textrm{eff}} =  \mathbb{M}_n \qty( \S_i \times \S_j) \cdot \vu n$ and thus $\mathbb{B}_4 = b_4 \mathbb{M}_w$ and $\mathbb{B}_5 = b_5 \mathbb{M}_v$. We observe that Hund's coupling is responsible for the anisotropy as $\mathbb{M}_u=\mathbb{M}_v=\mathbb{M}_w$ when $J_H/U \rightarrow 0$:
\begin{align}
\label{eq:Mconst}
  \lim_{J_H/U \rightarrow 0} \mathbb{M}_n = -\frac{16}{9} \frac{2t_1 + t_3 }{U} \equiv \mathbb{M}.
\end{align}
Unlike for Eq.~\eqref{eq:Aconstgeneral2}, $U \gg \lambda$ is not assumed. We see that in the limit $U \gg J_H$, the $\mathbb{M}_n$'s are independent of $\lambda$. The expressions for $\mathbb{M}_u$, $\mathbb{M}_v$, and $\mathbb{M}_w$ away from the $J_H/U \rightarrow 0$ limit are given in Appendix~\ref{sec:expression_edge}.

With the inclusion of the trigonal distortion at first order, we obtain additional terms. In the limiting case $U \gg J_H$,
\begin{align}
\label{eq:M_withCF}
 \lim_{\frac{J_H}{U} \rightarrow 0} \hat M_{u, \textrm{eff}}  = &\qty[\mathbb{M} - \frac{32}{81} \Delta \frac{2t_1 + 4t_2 + t_3 + 8t_4}{\lambda U}] \qty( \S_i \times \S_j)_u \nonumber \\
 \lim_{\frac{J_H}{U} \rightarrow 0} \hat M_{v, \textrm{eff}} = &\qty[\mathbb{M} + \frac{32}{81} \Delta \frac{2t_1 - 4t_2 + t_3 - 8t_4}{\lambda U}] \qty( \S_i \times \S_j)_v \nonumber \\
  &+ \qty[\sqrt{2} \frac{32}{81} \Delta \frac{2t_1 +t_3}{\lambda U}] \qty( \S_i \times \S_j)_w \nonumber \\
 \lim_{\frac{J_H}{U} \rightarrow 0} \hat M_{w , \textrm{eff}} = &\qty[\mathbb{M} - \frac{128}{81} \Delta \frac{t_2 + 2 t_4}{\lambda U}] \qty( \S_i \times \S_j)_w \nonumber \\
  &+ \qty[\sqrt{2}\frac{32}{81} \Delta \frac{2t_1 +t_3}{\lambda U}] \qty( \S_i \times \S_j)_v.
\end{align}
Equation~\eqref{eq:M_withCF} thus relates the coefficients $b_{1-5}$, obtained from the polarization integrals, to the coefficients $\mathbb{B}_i$ at first order in $\ham_{\textrm{hop}}$ and first order in $\Delta$. We note that the relations in Eq.~\eqref{eq:M_withCF} can be obtained without approximation numerically.
%The matrix elements in the ground state manifold can be calculated without approximation numerically by exact diagonalization of the two-site Hamiltonian followed by a projection into the magnetic Hilbert space (see Appendix for more details).
%Alternatively, the perturbation calculation exact in $\Delta$ can also be done numerically if we are only interested in the lowest order in $\ham_{\textrm{hop}}$.

Finally, we express $b_{1-5}$ in terms of the polarization integrals of Section~\ref{sec:polarization} using Eq.~\eqref{eq:phopp_eff}.
Without distortion, the ligands are in the $\vu x$ and $\vu y$ directions from the site $i$ with an atomic spacing $d=a/\sqrt{2}$ so that only $b_4$ and $b_5$ are finite,
\begin{align}
  b_4 &= - P^{\perp}_{d\pi d\delta} -\sqrt{2} \frac{P^{\parallel}_{pd\pi} t_{pd\pi}}{\Delta_{pd}} , \nonumber \\
  b_5 &=\frac{1}{4} \qty(\sqrt{6} P^{\perp}_{d\sigma d\pi} - \sqrt{2} P^{\perp}_{d\pi d\delta}) + \sqrt{2}\frac{P^{\perp}_{p\sigma d\pi} t_{pd\pi}}{\Delta_{pd}}.
  \label{eq:pol_int_b45}
\end{align}
Hence, even without the trigonal distortion, we recover the KNB formula ($b_4 = -b_5$) only when the polarization integrals are evaluated at zeroth order in $d$. In this case, the $d$-$d$ integrals vanish and $P^{\parallel}_{pd\pi}= P^{\perp}_{p\sigma d\pi}$. The results are summarized in Table~\ref{tab:results_edge}.

In Fig.~\ref{fig:numericsB}, we plot the coupling constants $\mathbb{B}_{2-4}$, calculated with exact diagonalization on a two-site cluster without any approximations, as a function of $J_H$, $\lambda$, and $\Delta$ in Figs.~\ref{fig:numericsB}(a), \ref{fig:numericsB}(b), and \ref{fig:numericsB}(c), respectively. Note that we cannot decrease $\lambda$ below a certain threshold in Fig.~\ref{fig:numericsB}(b) because the pseudospin 1/2 are not well defined for $\Delta <0$ when $\lambda = 0$ (the local ground state is fourfold degenerate). For simplicity, we assume $b_4 = -b_5$ and plotted the constants in units of $b_5$. The contributions $\mathbb{B}_{i}$ cannot be compared to the contributions $\mathbb{A}_i$ in Eq.~\eqref{eq:decomp} unless we accurately calculate the integrals in $b_4$ and $b_5$, which we do not attempt here. 

We see in both Figs.~\ref{fig:numericsA} and \ref{fig:numericsB} that the KNB formula (which corresponds to $m_4 = -m_5$ and $m_{1-3}=0$) is not accurate at all in the edge-sharing geometry, as $m_4 \neq m_5$ and all $m_{1-5}$ are finite in any parameter range.

The expressions taking into account the displacement of the ligands towards the bond ($\phi \neq 90^{\circ}$) is given in Appendix~\ref{sec:edge_angle}.

\subsection{Corner-sharing geometry}

\begin{table}[b]
\caption{Coupling coefficients defined in Eqs.~\eqref{eq:polcorner_allowed_CF} and \eqref{eq:decomp} for the corner-sharing geometry in the $J_H/U \rightarrow 0$ limit. The CF is included through $\ham_{\rm CF}$ but the displacement of the ligands is neglected}
\begin{tabular}{| c c c |}
  \hline
  $\mathbb{A}_1$ & $=$ & $0$ \\
  $\mathbb{A}_2$ & $=$ & $0$  \\
  \hline
  $\mathbb{B}_1$ & $=$ & $- t_{pd\pi} P^{\perp}_{p\pi d\sigma} \qty[\frac{32}{9}\frac{t}{U} - \frac{64}{81} \Delta \frac{t(U+6\lambda)} { U^2\lambda}]$ \\
  $\mathbb{B}_2$ & $=$ & $ t_{pd\pi} P^{\perp}_{p\pi d\sigma} \qty[\frac{32}{9}\frac{t}{U} +  \frac{128}{81} \Delta \frac{t(U -3\lambda)}{ U^2\lambda}]$ \\
  \hline
  \end{tabular}
\label{tab:results_corner}
\end{table}

%Here we proceed as in section \ref{sec:face_mech}, to which we refer for the various equations and some discussions.
In the corner-sharing geometry with $\varphi = 180^{\circ}$ [see Fig.~\ref{fig:geometries}(a)], the calculations are considerably simpler. The $\varphi \neq  180^{\circ}$ case is dealt with in Appendix~\ref{sec:corner_angle} and is more complex due to the lack of inversion symmetry.

\subsubsection{Lattice polarization}

We see from Eqs.~\eqref{eq:polcorner_allowed_noCF} and \eqref{eq:polcorner_allowed_CF} that polarization along the bond vanishes.
Because the only ligand is exactly at the center of the bond, its contribution also vanishes. Therefore, in contrast to the edge-sharing geometry, there is no ME coupling originating from the lattice polarization when $\varphi = 180^{\circ}$.

\subsubsection{Hopping polarization}
We define
\begin{align}
\label{eq:hop_N_op}
  \hat N_a = \vb c_i\dag  \mathcal{\hat A}_a \vb c_j + \textrm{H.c.},
\end{align}
in terms of the antisymmetric off-diagonal matrices $(\mathcal{\hat A}_a)_{bc} = \epsilon_{abc}$.
Then, for an $x$ bond, we find 
\begin{align}
  \vb P_{\textrm{hop}} = \qty[- t_{pd\pi} P^{\perp}_{p\pi d\sigma} \hat N_z] \vu y + \qty[t_{pd\pi} P^{\perp}_{p\pi d\sigma} \hat N_y] \vu z,
\end{align}

The expressions for other types of bonds are obtained by performing the proper rotations.

Casting the above operators in the low-energy subspace with perturbation theory at first order in $t$ and $\Delta$ in the limit $U \gg J_H$, we obtain
\begin{align}
\lim_{J_H/U \rightarrow 0} {\hat N}_{z, \textrm{eff}} & =\qty[\frac{32}{9}\frac{t}{U} - \frac{64}{81} \Delta \frac{t(U+6\lambda)} { U^2\lambda}]\qty( \vb S_i \times \vb S_j)_z \nonumber \\
\lim_{J_H/U \rightarrow 0}{\hat N}_{y, \textrm{eff}} &=\qty[\frac{32}{9}\frac{t}{U} +  \frac{128}{81} \Delta \frac{t(U -3\lambda)}{ U^2\lambda}]\qty( \vb S_i \times \vb S_j)_y.
 \end{align}
As for the edge-sharing geometry, $J_H$ is responsible for the anisotropy when $\Delta=0$. The results are summarized in Table~\ref{tab:results_corner}. The expressions away from the $J_H/U \rightarrow 0$ limit are given in Appendix~\ref{sec:expression_corner}.

We have thus obtained an expression for the two allowed ME coupling constants of Eq.~\eqref{eq:polcorner_allowed_CF} as a function of the various physical parameters and integrals in the $\varphi = 180^{\circ}$ geometry. The deviation from the KNB formula originating from the CF thus scales $\Delta/\lambda$ (for large $U$) relatively to the KNB $\Delta =0$ contribution. We stress that the scaling is only accurate in the $\lambda \gg \Delta$ limit.

%
%\subsection{Summary}
%
%In Table~\ref{tab:results_edge}, we summarized the resutls in the $U \gg J_H$ limit, where the CF is included through $\ham_{\rm CF}$ but the displacement of the ligands is neglected.
%
%

\section{Summary and conclusion}
\label{sec:discussion}

We developed a theory of the electric polarization in $d^5$ Mott insulators from electronic mechanisms. In particular we reconciled two approaches previously used to explain ME behaviors. On one hand, there is the ``hopping polarization'', which corresponds to formalisms principally used in the context of multiferroics, e.g. the KNB formula \cite{katsura2005spin} and other extensions \cite{miyahara2016theory}, and relies on matrix elements such as $\mel{d_{xy}}{y}{p_y}$. With an approach similar to Slater and Koster \cite{slater1954simplified}, we expanded the theory by taking into account the finite distance between ions and classified the different finite polarization integrals according to their symmetry.

On the other hand, the ``lattice polarization'' is intrinsic to the Hubbard model, merely defined by the positions of the electrons on the lattice. Charge effects in single-band Mott insulators due to such effects were first discussed in Ref.~\cite{bulaevskii2008electronic} and lead to the explanation of the subgap optical conductivity in different systems on lattices with triangular loops. In a previous work, we noted that in multiorbital systems with SOC, the same mechanism was possible at second order in the hopping \cite{bolens2018mechanism}. In the present paper, we thoroughly considered the phenomenon and conclude that (i) in the $\Delta_{pd} \gg U$ limit, the contribution from the positions of the ligand is suppressed by a factor of $U/\Delta_{pd}$ compared to the contributions form the TM ions, which itself vanishes if the trigonal distortion $\Delta \rightarrow 0$. Both contributions are relevant in the case of $\alpha$-RuCl$_3$ (see Fig.~\ref{fig:numericsA}). We also conclude that (ii) the edge-sharing geometry is crucial to obtain a finite polarization operator along the TM-TM bond. In the corner-sharing geometry, even when $\varphi \neq 180^{\circ}$ (see Appendix~\ref{sec:corner_angle}), the polarization along the bond vanishes.

%The two mechanisms are complementary.
%, even though the direction of the lattice polarization is restricted by the positions of the ions.
The relative contributions of the two mechanisms are determined by comparing the polarization integral $p_{\rm eff}$ with $a \cdot t_{\rm eff}/U$ between TM ions separated by a distance $a$. Together with complementary first-principle methods to evaluate the different physical parameters, in particular the polarization integrals, the ME effects in $d^5$ Mott insulators can be predicted from our results. In particular, in the edge-sharing geometry the direction of the polarization changes whether $p_{\rm eff} \ll a\cdot t_{\rm eff}/U$, in which case $\mathbb{A}_1$ is finite (see Table~\ref{tab:results_edge}) and $\vb P$ has a contribution along the bond, or $p_{\rm eff} \gg a\cdot t_{\rm eff}/U$, in which case $\vb P$ is perpendicular to the bond. In any case, we generally find significant deviations from the KNB formula in the edge-sharing geometry, as can be seen in Figs.~\ref{fig:numericsA} and \ref{fig:numericsB}.
%Finally, we note that the different spin-polarization coupling constants in Eq.~\eqref{eq:decomp} can be numerically evaluated more accurately by performing the perturbation theory calculations numerically or by using exact diagonalization.

We expect that the THz optical conductivity observed in $\alpha$-RuCl$_3$ \cite{little2017antiferromagnetic, wang2017magnetic, wellm2017signatures, reschke2018sub, shi2018field} can be better understood using our results by calculating the dynamical response of the effective polarization operator in the magnetic ground state manifold, which has only been done in the pure Kitaev model \cite{bolens2018mechanism}. 

From our results in the edge-sharing geometry of Kitaev materials, the intensities of the regular magnetic dipole-induced absorption and the electric dipole-induced absorption can be compared. 
Without calculating the correlation functions, the ratio of the electric dipole contribution to the magnetic dipole contribution can be estimated as $(|\hat m|/g\lambdabar)^2$, where $|\hat m|$ is of the order of the spin-polarization coupling constants $\{m_i\}$, $g$ is the effective Land\'e $g$ factor, and $\lambdabar$ is the Compton wavelength \cite{bolens2018mechanism}. From the `lattice polarization' contributions (see Table~\ref{tab:results_edge} and Fig.~\ref{fig:numericsA}), we find that the electric dipole-induced absorption is around $20$ times larger than the magnetic dipole-induced one. The `hopping polarization' contribution could be even larger but cannot be estimated without calculating the integrals in Eq.~\eqref{eq:pol_int_b45}. 
Finally, the electric dipole-induced absorption is expected for both in-plane and out-of-plane polarizations with similar amplitudes.

\section{Acknowledgments}
A. B. thanks H. Katsura for fruitful discussions and acknowledges the Leading Graduate Course for Frontiers of Mathematical Sciences and Physics (FMSP) for the encouragement of the present paper.

\appendix

\section{Edge-sharing geometry with $\phi \neq 90^{\circ}$}
\label{sec:edge_angle}

Here, we consider the effect of the distortion corresponding to the two ligands moving toward the center of the bond in the $\pm \vu v$ directions.
As mentioned in Sec.~\ref{sec:edge}, the displacement of the ligands is along one of the $C_{2}$ axis so that the processes will only yield coupling consistent with the $D_{2h}$ symmetry group. This lets us study the angle dependence away from the ideal $\phi=90^{\circ}$ bond geometry.

The hopping integrals can be expressed via the Slater-Koster integrals ($t_{pd\sigma}, t_{pd\pi}, t_{dd\sigma}, t_{dd\pi}$, and $t_{dd\delta}$),
\begin{align}
t_1 &= \frac{t_{dd\pi} + t_{dd\delta}}{2}  + \frac{t_{pd\pi}^2}{\Delta_{pd}}\cos(\phi) \nonumber \\ 
t_2 &= \frac{- t_{dd\pi} + t_{dd\delta}}{2} + \frac{t_{pd\pi}^2}{\Delta_{pd}} \nonumber \\
t_3 &= \frac{3t_{dd\sigma} + t_{dd\delta}}{4} + \frac{3}{2} \frac{t_{pd\sigma}^2}{\Delta_{pd}} \cos^3(\phi)\nonumber \\
 &+ \frac{ \qty(\sqrt{3} t_{pd\sigma} - t_{pd\pi})t_{pd\pi}}{\Delta_{pd}} \sin(\phi)\sin(2\phi).
\end{align}
To be precise, we note that the integrals themselves also depend on $\phi$ due to the change in the TM-L distance (or alternatively in the TM-TM distance). We do not explicitly consider this dependence.\footnote{Our results are similar to those of Ref.~\cite{winter2016challenges} except for $t_3$, which might be caused by a different microscopic definition of $\phi$.}

\subsection{Mechanism}
\subsubsection{Lattice polarization}
The two ligands are separated by a distance $\tilde a = a \cot(\phi/2)$ in the $\vu v$ direction. The lattice polarization expression for the ligands, Eq.~\eqref{eq:pollat_ligand} when $\phi = 90^{\circ}$, becomes
\begin{equation}
\vb P_{\textrm{lat}}^{\textrm{(L)}} = \frac{t_{pd\pi}^2}{\Delta_{pd}^2} \sin(\phi) \qty[ \vb c\dag_i
  \begin{pmatrix}
  	0 & -1 & 0 & \\
  	1 & 0 & 0 \\
  	0 & 0 & 0 \\
  \end{pmatrix}
  \vb c_j + \textrm{H.c.} ] \frac{\tilde a}{2} \vu v.
\end{equation}
Here the $t_{pd\pi}$ integral also depends on $\phi$ if the TM-L distance $d= a/(2\cos(\phi/2))$ varies.

\subsubsection{Lattice polarization}
As in the $\phi=90^{\circ}$ situation, only $b_4$ and $b_5$ are nonzero without the trigonal distortion. We find,
\begin{align}
 b_4 &= - P^{\perp}_{d\pi d\delta} - 2\sin(\phi/2) \frac{t_{pd\pi}}{\Delta_{pd}} \qty(
  P^{\parallel}_{pd\pi} + \cos(\phi)(P^{\parallel}_{pd\pi} - P^{\perp}_{p\pi d\delta})) \nonumber \\
 b_5 &= \frac{1}{4} \qty(\sqrt{6} P^{\perp}_{d\sigma d\pi} - \sqrt{2} P^{\perp}_{d\pi d\delta}) + \frac{\sin(\phi/2)}{2} \frac{1}{\Delta_{pd}} \times  \\
   &\Big[ \cos(\phi ) \left(2
   \sqrt{3}P^{\perp}_{p\pi d\sigma} t_{pd\pi} + P^{\perp}_{p\pi d\delta}(- 2 \sqrt{3}t_{pd\sigma}+6t_{pd\pi})\right)
    \nonumber \\
   &+ \cos(2 \phi )( P^{\perp}_{p\sigma d\pi} - P^{\perp}_{p\pi d\delta}) (\sqrt{3} t_{pd\sigma}-2 t_{pd\pi} )
  \nonumber \\
   &+P^{\perp}_{p\sigma d\pi}( \sqrt{3} t_{pd\sigma}+ 2 t_{pd\pi}) + P^{\perp}_{p\pi d\delta}( - \sqrt{3}
   t_{pd\sigma} +  2 t_{pd\pi}) \Big] \nonumber.
\end{align}

With the addition of the trigonal distortion (displacement of the ligands perpendicular to the plane), all $b_{1-5}$ are finite and the calculation is too cumbersome, but can be done numerically.

\section{Corner-sharing geometry with $\varphi \neq 180^{\circ}$}
\label{sec:corner_angle}

In this appendix, we consider the more general corner-shared geometry where a relative rotation along the $\vu z$ axis is allowed such that $\varphi\neq 180$ [see Fig.~\ref{fig:geometries}(a)]. In this case, there is no inversion symmetry at the center of the bond. The cubic orbitals are defined relatively to two sets of unit vectors $\mathcal{B}_A = \{\vu x_A, \vu y_A, \vu z\}$ and $\mathcal{B}_B = \{\vu x_B, \vu y_B, \vu z\}$ separated by a $2 \alpha = \pi - \varphi$ rotation along the $\vu z$ axis. We consider a uniform tetragonal distortion $\Delta_A = \Delta_B = \Delta$. A staggered tetragonal distortion (corresponding to $\Delta_A = -\Delta_B$) is also plausible, as shown in Ref.~\cite{torchinsky2015structural}, but we do not consider it explicitly. For an $x$ bond, we define $\alpha>0$ for a displacement of the shared ligand in the $+\vu y$ direction.
\subsubsection{Full octahedral symmetry with $\alpha \neq 0$}

For $\alpha \neq 0$, the symmetry group of the bond is reduced to $C_{2v}$. Its character table is shown in Table \ref{tab:C2v}. Importantly, there is no inversion symmetry at the center of the bond. Thus, the hopping can be antisymmetric and the polarization can couple to symmetric two-spin operators. The full ME coupling is then characterized by at most thirty coefficients instead of nine.

The effective hopping matrices between the $t_{2g}$ orbitals of site $i$ and $j$ are
\begin{align}
\label{eq:hop_C2v}
  \hat T_x(C_{2v}) = 
  \begin{pmatrix}
    t_{1} & t_4 & 0 \\
	-t_4 & t_{2} & 0 \\
	0 & 0 & t_{3}
  \end{pmatrix} \qq{and}
\hat T_y(C_{2v}) = 
  \begin{pmatrix}
    t_{2} & t_4 & 0 \\
	-t_4 & t_{1} & 0 \\
	0 & 0 & t_{3}
  \end{pmatrix}.
\end{align}
Considering only indirect $p$-$d$ hopping and working in the $\mathcal{B}_A$ and $\mathcal{B}_B$ bases, we have simply
\begin{align}
  \hat T_x &= 
  \begin{pmatrix}
    0 & 0 & 0 \\
	0 & t_2 & 0 \\
	0 & 0 & t_2\cos(2\alpha)
  \end{pmatrix}_{\mathcal{B}_A, \mathcal{B}_B}\qq{and} \\
   \hat T_y &= 
  \begin{pmatrix}
    t_2 & 0 & 0 \\
	0 & 0 & 0 \\
	0 & 0 & t_2\cos(2\alpha)
  \end{pmatrix}_{\mathcal{B}_A, \mathcal{B}_B}.
\end{align}
 
The ME coupling (for an $x$ bond) becomes 
\begin{align}
  \label{eq:coupling_withoutinv}
%  \begin{pmatrix}
%  	P_x \\
%  	P_y \\
%  	P_z
%  \end{pmatrix}
\vb P
   =& 
  \begin{pmatrix}
    0 & 0 & 0 \\
	0 & 0 & m_1^A \\
	0 & m_2^A & 0
  \end{pmatrix}
  \qty(\vb S_i \times \vb S_j) \nonumber \\
  &+ \begin{pmatrix}
  	0 & 0 & 0 & 0 \\
  	m_0^D & m_1^D & m_2^D & m_3^D \\
  	0 & 0 & 0 & 0
  \end{pmatrix}
  \begin{pmatrix}
  	\mathbb{I} \\ S_i^x S_j^x \\ S_i^y S_j^y \\ S_i^z S_j^z
  \end{pmatrix} \nonumber \\
   &+ \begin{pmatrix}
    0 & 0 & m_1^{\Gamma} \\
	0 & 0 & 0 \\
	m_2^{\Gamma} & 0 & 0
  \end{pmatrix}
  \begin{pmatrix}
    S_i^y S_j^z + S_i^z S_j^y \\
	S_i^x S_j^z + S_i^z S_j^x \\
	S_i^x S_j^y + S_i^y S_j^x
  \end{pmatrix}.
\end{align}

Spin operators in the $\mathcal{B}_{A(B)}$ basis are obtained by the transformation $\vb S_i \rightarrow e^{i \alpha S^z} \vb S_i  e^{-i \alpha S^z}$ ($\vb S_j \rightarrow e^{-i \alpha S^z} \vb S_j  e^{i \alpha S^z}$).
When considering the two-spin operators, the change of basis $(\mathcal{B}, \mathcal{B}) \rightarrow (\mathcal{B}_A, \mathcal{B}_B)$ does not change the representation of the operators, so that Eq.~\eqref{eq:coupling_withoutinv} is valid with spin operators written in either set of bases (the values of the coefficients are nevertheless modified). From now on we work with the $\mathcal{B}$ basis for $\vb P$ and in the $\mathcal{B}_A$ and $\mathcal{B}_B$ bases for $\vb S_i$ and $\vb S_j$, respectively.

\subsubsection{Tetragonal distortion}  
For a uniform tetragonal distortion, the $C_{2v}$ symmetry group is not affected by $\Delta$ and the form of $\hat m$ does not change.
%(For a staggered tetragonal distortion though, the symmetry is reduced to $C_s$ when $\alpha \neq 0$)

\subsection{Mechanism}
\subsubsection{Lattice polarization}

We see from Eq.~\eqref{eq:coupling_withoutinv} that polarization along the bond is only possible via $m_1^\Gamma$. However, as discussed in Sec.~\ref{sec:face_mech}, the contribution from the TM ions must vanish when the hopping is diagonal with respect to the pseudospins, which happens when the $3\times 3$ matrices defining the tetragonal CF and the hopping commute. In the present case, due to the forms of $\ham_{\textrm{CF}}$ and \eqref{eq:hop_C2v}, both matrices always commute even in the lowest symmetry, so that $\vb P_{\textrm{lat}}^{\textrm{(TM)}} = 0$.

The ligand contribution corresponds exactly to the hopping matrix (neglecting $d$-$d$ direct hopping),
\begin{align}
 \label{eq:pollat180_ligand}
\vb P_{\textrm{lat}, x}^{\textrm{(L)}} &= \tan(\alpha) \qty[ \vb c\dag_{i}
 \hat T_{x}
  \vb c_{j} + \textrm{H.c.} ]\frac{a}{2}\vu y \\
  \vb P_{\textrm{lat},y}^{\textrm{(L)}} &= -\tan(\alpha) \qty[ \vb c\dag_{i}
 \hat T_{y}
  \vb c_{j} + \textrm{H.c.} ]\frac{a}{2}\vu x.
\end{align}
We show in the next subsection that such microscopic operators will, in turn, only produce a coupling with diagonal two-spin operators $S_i^{\alpha}S_j^{\alpha}$ in addition to a spin-independent polarization (expected due to the lack of inversion symmetry). 

\subsubsection{Hopping polarization}
For an $x$ bond, we find 
\begin{align}
  \vb P_{\textrm{hop}} = \qty[\hat N^x_S + \hat N^x_A] \vu x  + \qty[\hat N^y_S + \hat N^y_A] \vu y + \qty[\hat N^z_S + \hat N^z_A] \vu z,
\end{align}
with
\begin{align}
  \hat N^a_{S/A} = \vb c_i\dag  \mathcal{\hat N}^a_{S/A} \vb c_j + \textrm{H.c.},
\end{align}
%\begin{align}
%  \vb P_{\textrm{hop}} &= \qty[
%   \vb c_i\dag \qty( \mathbb{S}^x_{S} + \mathbb{S}^x_{A} ) \vb c_j + \textrm{H.c.},] \vu x \nonumber \\
%   &+ \qty[
%   \vb c_i\dag \qty( \mathbb{S}^y_{S} + \mathbb{S}^y_{A} ) \vb c_j + \textrm{H.c.},] \vu y  \nonumber \\
%   &+ \qty[
%   \vb c_i\dag \qty( \mathbb{S}^z_{S} + \mathbb{S}^z_{A} ) \vb c_j + \textrm{H.c.},] \vu z,
%\end{align}
where $S$ and $A$ stand for symmetric and antisymmetric, which are both allowed when $\alpha \ne 0$ due to the lack of inversion symmetry at the center of the bonds. Denoting the antisymmetric and symmetric off-diagonal matrices by $(\mathcal{\hat A}_a)_{bc} = \epsilon_{abc}$ and $(\mathcal{\hat S}_a)_{bc} = \abs{\epsilon_{abc}}$, respectively ($\epsilon$ is the Levi-Civita tensor), we have
\begin{align}
\label{eq:polhopcorner_matrices}
\mathcal{\hat N}^x_{S}&=  \sin(\alpha) t_{pd\pi} P^{\perp}_{p\pi d\sigma}
%\begin{pmatrix}
%  	0 &  1 & 0 & \\
%  	1 & 0 & 0 \\
%  	0 & 0 & 0 
%  \end{pmatrix}
\mathcal{\hat S}_z
  , \quad \mathcal{\hat N}^x_A = 0,\nonumber \\
\mathcal{\hat N}^y_{S} &=  -2 \sin(\alpha) t_{pd\pi} \times \nonumber \\
 &
 \begin{pmatrix}
  	0 & & 0 & \\
   0 & P^{\parallel}_{pd\pi} & 0 \\
  	0 & 0 & P^{\perp}_{p\sigma d\pi} + (P^{\parallel}_{pd\pi} + P^{\perp}_{p\sigma d\pi} ) \cos(2\alpha)
\end{pmatrix}
,\nonumber \\
  \mathcal{\hat N}^y_{A}&=  -\cos(\alpha) t_{pd\pi} P^{\perp}_{p\pi d\sigma} 
%  \begin{pmatrix}
%  	0 &  1 & 0 & \\
%  	- 1 & 0 & 0 \\
%  	0 & 0 & 0 
%  \end{pmatrix}
  \mathcal{\hat A}_z,\nonumber \\
\mathcal{\hat N}^z_{S}&=  -\sin(2\alpha) t_{pd\pi}P^{\perp}_{p\sigma d\pi} 
%\begin{pmatrix}
%  	0 &  0 & 0 & \\
%  	0 & 0 & 1 \\
%  	0 & 1 & 0 
%  \end{pmatrix}
  \mathcal{\hat S}_x, \nonumber \\
\mathcal{\hat N}^z_{A}&=  \cos(2\alpha) t_{pd\pi} P^{\perp}_{p\pi d\sigma} 
%\begin{pmatrix}
%  	0 &  0 & 1 & \\
%  	0 & 0 & 0 \\
%  	-1 & 0 & 0 
%  \end{pmatrix}
  \mathcal{\hat A}_y.
\end{align}
The expressions for a $y$ bond are obtained by performing a $\pi/2$ rotation around $\vu z$.

Casting the above operators in the low-energy subspace, we obtained the effective polarization operator. First, even in the $C_{2v}$, we can read from the character tables that $\mathcal{\hat A}_{a, \textrm{eff}} = \mathfrak{a}_a \qty( \vb S_i \times \vb S_j)_a$ and $\mathcal{\hat S}_{a, \textrm{eff}} = \mathfrak{s}_a \abs{\epsilon_{abc}} S_i^b S_j^c$, respectively. For the antisymmetric part, the results from perturbation theory in the limit $U \gg J_H$ are
\begin{align}
  \mathfrak{a}_z &= \frac{32}{9}\frac{t_2\cos^2(\alpha)}{U} + \Delta \frac{64}{81} \frac{t_2 (-2U-3\lambda +(U-3\lambda)\cos(2\alpha))} { U^2\lambda} \nonumber \\
  \mathfrak{a}_y &= \frac{32}{9}\frac{t_2\cos^2(\alpha)}{U} + \Delta \frac{32}{81} \frac{t_2 (-U-6\lambda +(5U-6\lambda)\cos(2\alpha))} { U^2\lambda}.
 \end{align}
As for the edge-sharing geometry, we observe that $J_H$ is responsible for the anisotropy when $\Delta=0$. For the symmetric part, the off-diagonal matrices give a contribution that is proportional to $J_H$. In the $U \gg J_H$ limit,
\begin{align}
  \mathfrak{s}_z =& \frac{32}{9}\frac{t_2J_H(1-2\cos(2\alpha))}{(2U+3\lambda)^2} \nonumber \\
  & - \Delta \frac{128}{81}\frac{t_2 J_H (2U-9\lambda) (-1 +2\cos(2\alpha)) } { \lambda (2U+3\lambda)^2} \nonumber \\
  \mathfrak{s}_x =& \frac{64}{9}\frac{t_2J_H\cos^2(\alpha)}{(2U+3\lambda)^2} \nonumber \\
  & - \Delta \frac{128}{81}\frac{t_2 J_H (U+18\lambda + (U+9\lambda)\cos(2\alpha)) } { \lambda (2U+3\lambda)^2}.
 \end{align}
Finally, the diagonal terms in Eq.~\eqref{eq:polhopcorner_matrices} each give a constant and three $S_i^a S_j^a$ terms with $a=x$, $y$, and $z$. Moreover, in the absence of $J_H$, the contribution is isotropic. In the $U \gg J_H$ and $U \gg \lambda$ limit,
\begin{align}
\begin{pmatrix}
  	0 & 0 & 0 \\
  	0 & 1 & 0 \\
  	0 & 0 & 0 
\end{pmatrix}& \rightarrow
\qty[\frac{16}{9} \frac{t_2 \cos^2(\alpha)}{U} + \Delta \frac{32}{81} \frac{t_2 (-2 + \cos(2\alpha))}{\lambda U} ]\vb S_i \cdot \vb S_j \nonumber \\
 &+ \frac{8}{9} \frac{t_2(-5 + \cos(2\alpha))}{U} + \Delta \frac{32}{81} \frac{t_2 (4 + \cos(2\alpha))}{\lambda U}, \nonumber \\
 \begin{pmatrix}
  	0 & 0 & 0 \\
  	0 & 0 & 0 \\
  	0 & 0 & 1 
\end{pmatrix}& \rightarrow
\qty[\frac{16}{9} \frac{t_2 \cos^2(\alpha)}{U} + \Delta \frac{32}{81} \frac{t_2 (1 + 4\cos(2\alpha))}{\lambda U} ]\vb S_i \cdot \vb S_j \nonumber \\
 &+ \frac{8}{9} \frac{t_2(1 -5 \cos(2\alpha))}{U} + \Delta \frac{32}{81} \frac{t_2 (1 -8 \cos(2\alpha))}{\lambda U}.
\end{align}

We have thus obtained an expression for all the ME coupling constants of Eq.~\eqref{eq:coupling_withoutinv} as a function of the various physical parameters and integrals.

\section{Character tables}
\label{sec:tables}
Here we show the character \cref{tab:D2,tab:C2,tab:D4,tab:C2v} and the different functions corresponding to each irreducible representations for the symmetry groups mentioned throughout the paper.

\begin{table*}[h]
\centering
\caption{Character table of $D_{2h}$ with twofold axis in the $a$, $b$, and $c$ directions.
% with main axes $\vu x$, $\vu y$ and $\vu z$. 
$C_{2, \alpha}$ is a $C_2$ rotation along the corresponding axis ($\alpha = a,b,$ or $c$), $\sigma_{\alpha\beta}$ is a reflection across the $\alpha\beta$ plane, and the $\langle ij \rangle$ bond is in the $a$ direction.}
\begin{tabular}{| c " c | c | c | c | c | c | c | c " c |}
  \hline
  $D_{2h}$ & $E$ & $C_{2,a}$ & $C_{2,b}$ & $C_{2,c}$ & $I$ & $\sigma_{bc}$ & $\sigma_{ac}$ & $\sigma_{ab}$ & functions  \\
  \hline
  \hline
  $A_g$ & $1$ & $1$ & $1$ & $1$ & $1$ & $1$ & $1$ & $1$ & -\\
  \hline
  $B_{1,g}$ & $1$ & $1$ & $-1$ & $-1$ & $1$ & $1$ & $-1$ & $-1$ & $S^a$\\
  \hline
  $B_{2,g}$ & $1$ & $-1$ & $1$ & $-1$ & $1$ & $-1$ & $1$ & $-1$ & $S^b$\\
  \hline
  $B_{3,g}$ & $1$ & $-1$ & $-1$ & $1$ & $1$ & $-1$ & $-1$ & $1$ & $S^c$\\
  \hline \hline
  $A_u$ & $1$ & $1$ & $1$ & $1$ & $-1$ & $-1$ & $-1$ & $-1$ &$\qty(\vb S_i \times \vb S_j)_a$\\
  \hline
  $B_{1,u}$ & $1$ & $1$ & $-1$ & $-1$ & $-1$ & $-1$ & $1$ & $1$ & $a$, $i$-$j$ antisym.  \\
%  &&&&&&&&&$a_ib_{j} \rightarrow a_ib_j - a_jb_i$\\
  \hline
  $B_{2,u}$ & $1$ & $-1$ & $1$ & $-1$ & $-1$ & $1$ & $-1$ & $1$& $b$, $\qty(\vb S_i \times \vb S_j)_b$\\
  \hline
  $B_{3,u}$ & $1$ & $-1$ & $-1$ & $1$ & $-1$ & $1$ & $1$ & $-1$ & $c$, $\qty(\vb S_i \times \vb S_j)_c$\\
  \hline
  \end{tabular}
\label{tab:D2}
\end{table*}

\begin{table*}[h]
\vspace{5pt}
\centering
\caption{Character table of $C_{2h}$ with twofold axis in the $a$ direction.
% with main axes $\vu x$, $\vu y$ and $\vu z$
$C_{2,a}$ is the $C_2$ rotation around $\vu a$, $\sigma_{bc}$ is the reflection across the plane perpendicular to $\vu a$, and the $\langle ij \rangle$ bond is along the $a$ direction.}
\begin{tabular}{| c " c | c | c | c | c | c " c |}
  \hline
  $C_{2h}$ & $E$ & $C_{2,a}$ & $I$ & $\sigma_{bc}$& functions  \\
  \hline
  \hline
  $A_{g}$ & $1$ & $1$ & $1$ & $1$ & $S^a$ \\
  \hline
  $B_{g}$ & $1$ & $-1$ & $1$ & $-1$ &$S^b$, $S^c$\\
  \hline \hline
  $A_u$ & $1$ & $1$ & $-1$ & $-1$ & $a$, $\qty(\vb S_i \times \vb S_j)_a$, $i$-$j$ antisym.  \\
  \hline
  $B_{u}$ & $1$ & $-1$ & $-1$ & $1$ & $b$, $\qty(\vb S_i \times \vb S_j)_b$ \\
  &&&&& $c$, $\qty(\vb S_i \times \vb S_j)_c$  \\
  \hline
  \end{tabular}
\label{tab:C2}
\end{table*}

\begin{table*}[h]
\caption{Character table of $D_{4h}$ with $\vu a$ as the $C_4$ axis. The $2C'_2$ rotations are around $\vu b$ and $\vu c$, the $2C''_2$ ones around $\vu b \pm \vu c$, and $2 \sigma_{v,d}$ are the corresponding reflections. The $\langle ij \rangle$ bond is along $\vu a$.}
\begin{tabular}{| c " c | c | c | c | c | c | c | c | c | c " c |}
  \hline
  $D_{4h}$ & $E$ & $2C_{4,a}$ & $C_{2,a}$ & $2C'_{2}$ & $2C''_{2}$ & $I$ & $2S_{4}$ & $\sigma_{bc}$ & $2\sigma_{v}$ & $2\sigma_{d}$ & functions  \\
  \hline
  \hline
  $A_{1,g}$ & $1$ & $1$ & $1$ & $1$ & $1$ & $1$ & $1$ & $1$ & $1$ & $1$ & -\\
  \hline
  $A_{2,g}$ & $1$ & $1$ & $1$ & $-1$ & $-1$ & $1$ & $1$ & $1$ & $-1$ & $-1$ &$S^a$ \\
  \hline
  $B_{1,g}$ & $1$ & $-1$ & $1$ & $1$ & $-1$ & $1$ & $-1$ & $1$ & $1$ & $-1$ & -\\
  \hline
  $B_{2,g}$ & $1$ & $-1$ & $1$ & $-1$ & $1$ & $1$ & $-1$ & $1$ & $-1$ & $1$ & $bc$\\
  \hline
  $E_{g}$ & $2$ & $0$ & $-2$ & $0$ & $0$ & $2$ & $0$ & $-2$ & $0$ & $0$ & $(S^b, S^c)$, $(ac, ab)$ \\
  \hline \hline
   $A_{1,u}$ & $1$ & $1$ & $1$ & $1$ & $1$ & $-1$ & $-1$ & $-1$ & $-1$ & $-1$ & -\\
  \hline
  $A_{2,u}$ & $1$ & $1$ & $1$ & $-1$ & $-1$ & $-1$ & $-1$ & $-1$ & $1$ & $1$ &$a$, $i$-$j$ antisym. \\
  \hline
  $B_{1,u}$ & $1$ & $-1$ & $1$ & $1$ & $-1$ & $-1$ & $1$ & $-1$ & $-1$ & $1$ & $(\vb S_i \times \vb S_j)_a$\\
  \hline
  $B_{2,u}$ & $1$ & $-1$ & $1$ & $-1$ & $1$ & $-1$ & $1$ & $-1$ & $1$ & $-1$ & -\\
  \hline
  $E_{u}$ & $2$ & $0$ & $-2$ & $0$ & $0$ & $-2$ & $0$ & $2$ & $0$ & $0$ & $(b, c)$, \\
  &&&&&&&&&&& $\qty((\vb S_i \times \vb S_j)_b, (\vb S_i \times \vb S_j)_c)$ \\
  \hline 
%  
%  $A_u$ & $1$ & $1$ & $1$ & $1$ & $-1$ & $-1$ & $-1$ & $-1$ & & &$\qty(\vb S_i \times \vb S_j) \cdot \vu u$\\
%  \hline
%  $B_{1,u}$ & $1$ & $1$ & $-1$ & $-1$ & $-1$ & $-1$ & $1$ & $1$ & & & $\vb r\cdot \vu u$,  \\
%%  &&&&&&&&&$a_ib_{j} \rightarrow a_ib_j - a_jb_i$\\
%  \hline
%  $B_{2,u}$ & $1$ & $-1$ & $1$ & $-1$ & $-1$ & $1$ & $-1$ & $1$ & & &  $\vb r\cdot \vu v$, $\qty(\vb S_i \times \vb S_j) \cdot \vu w$\\
%  \hline
%  $B_{3,u}$ & $1$ & $-1$ & $-1$ & $1$ & $-1$ & $1$ & $1$ & $-1$ & & & $\vb r\cdot \vu w$, $\qty(\vb S_i \times \vb S_j) \cdot \vu v$\\
%  \hline
 \end{tabular}
\label{tab:D4}
\end{table*}

\begin{table*}[h]
\centering
\caption{Character table of $C_{2v}$. The $\langle ij \rangle$ bond is along the $\vu x$ axis.}
\begin{tabular}{| c " c | c | c | c " c |}
  \hline
  $C_{2v}$ & $E$ & $C_{2,y}$ & $\sigma_{xy}$ & $\sigma_{yz}$ & functions  \\
  \hline
  \hline
  $A_{1}$ & $1$ & $1$ & $1$ & $1$ &  $y$, $\qty(\vb S_i \times \vb S_j)_z$\\
  \hline
  $A_{2}$ & $1$ & $1$ & $-1$ & $-1$ &  $xz$, $S^y$, $\qty(\vb S_i \times \vb S_j)_x$\\
  \hline
  $B_{1}$ & $1$ & $-1$ & $1$ & $-1$ & $x$, $xy$, $S^y$, $i$-$j$ antisym.\\
  \hline
  $B_{2}$ & $1$ & $-1$ & $-1$ & $1$ & $z$, $yz$, $S^y$, $\qty(\vb S_i \times \vb S_j)_y$\\
  \hline
  \end{tabular}
\label{tab:C2v}
\end{table*}
\clearpage

\onecolumngrid
\section{Polarization integrals}
\label{sec:polApendix}
In Table~\ref{tab:pol} we list the polarization integrals for a general bond direction in a manner similar to that of Ref.~\cite{slater1954simplified}. The entries not given can be found by cyclically permuting the coordinates and direction cosines.
\begin{table}[h]
\caption{Polarization integrals in terms of two-center integrals for two ions separated by a vector $\vb d = \norm{\vb d}(l,m,n)$. Only the $p$-$t_{2g}$ and $t_{2g}$-$t_{2g}$ integrals are considered.}
\begin{tabular}{| c c c |}
  \hline
  $\mel{p_{x}}{x}{d_{yz}}$ & $=$ &
  $
  \sqrt{3} l^2 m n                 P^{\parallel}_{pd\sigma}
- 2  l^2 m n 		            P^{\parallel}_{pd\pi}
+ \sqrt{3} m n \qty(1-l^2) 		P^{\perp}_{p\pi d\sigma}
- 2 l^2 m n 			        P^{\perp}_{p\sigma d\pi}
+ m n  \qty(1+l^2)         P^{\perp}_{p\pi d\delta}$ \\

  $\mel{p_{x}}{y}{d_{yz}}$ & $=$ &
  $
  \sqrt{3} l m^2 n               P^{\parallel}_{pd\sigma}
- 2  l m^2 n               P^{\parallel}_{pd\pi}
- \sqrt{3} l m^2	 n			 P^{\perp}_{p\pi d\sigma}
+l n \qty(1 - 2 m^2)  			   P^{\perp}_{p\sigma d\pi}
- l n \qty(1-m^2)        P^{\perp}_{p\pi d\delta}$ \\

  $\mel{p_{x}}{z}{d_{xy}}$ & $=$ &
  $
  \sqrt{3} l m n^2               P^{\parallel}_{pd\sigma}
- 2  l m n^2               P^{\parallel}_{pd\pi}
- \sqrt{3} l m n^2			 P^{\perp}_{p\pi d\sigma}
+l m \qty(1 - 2 n^2)  			   P^{\perp}_{p\sigma d\pi}
- l m \qty(1-n^2) 		        P^{\perp}_{p\pi d\delta}$ \\
  \hline

  \hline
  $\mel{p_{x}}{x}{d_{xz}}$ & $=$ &
  $
  \sqrt{3} l^3 n                 P^{\parallel}_{pd\sigma}
+  l n \qty(1-2l^2)		            P^{\parallel}_{pd\pi}
+ \sqrt{3} l n \qty(1-l^2) 		P^{\perp}_{p\pi d\sigma}
+ l n \qty(1 - 2l^2) 	        P^{\perp}_{p\sigma d\pi}
- l n  \qty(1-l^2)         P^{\perp}_{p\pi d\delta}$ \\

  $\mel{p_{x}}{y}{d_{xz}}$ & $=$ &
  $
  \sqrt{3} l^2 m n               P^{\parallel}_{pd\sigma}
+ m n \qty( 1 - 2l^2)               P^{\parallel}_{pd\pi}
- \sqrt{3} l^2 m	 n			 P^{\perp}_{p\pi d\sigma}
- 2 l^2 m n			  	   P^{\perp}_{p\sigma d\pi}
- m n \qty(1-l^2)      	  P^{\perp}_{p\pi d\delta}$ \\

  $\mel{p_{x}}{z}{d_{xz}}$ & $=$ &
  $
  \sqrt{3} l^2 n^2               P^{\parallel}_{pd\sigma}
+ n^2 \qty( 1 - 2l^2)               P^{\parallel}_{pd\pi}
- \sqrt{3} l^2 n^2			 P^{\perp}_{p\pi d\sigma}
+ l^2 \qty(1 - 2 n^2)  			   P^{\perp}_{p\sigma d\pi}
+ \qty(1-l^2) \qty(1-n^2)        P^{\perp}_{p\pi d\delta}$ \\
  \hline

  \hline
  $\mel{p_{x}}{x}{d_{xy}}$ & $=$ &
  $
  \sqrt{3} l^3 m                 P^{\parallel}_{pd\sigma}
+ l m \qty(1-2l^2)               P^{\parallel}_{pd\pi}
+ \sqrt{3} lm \left(1-l^2\right) P^{\perp}_{p\pi d\sigma}
+ l m \left(1-2l^2\right)        P^{\perp}_{p\sigma d\pi}
- l m \left(1-l^2\right)         P^{\perp}_{p\pi d\delta}$ \\

  $\mel{p_{x}}{y}{d_{xy}}$ & $=$ &
  $
  \sqrt{3} l^2 m^2               P^{\parallel}_{pd\sigma}
+ m^2 \qty(1-2l^2)               P^{\parallel}_{pd\pi}
- \sqrt{3} l^2 m^2				 P^{\perp}_{p\pi d\sigma}
+l^2 \qty(1 - 2 m^2)   			   P^{\perp}_{p\sigma d\pi}
+\qty(1-l^2) \qty(1-m^2)        P^{\perp}_{p\pi d\delta}$ \\

  $\mel{p_{x}}{z}{d_{xy}}$ & $=$ &
  $
  \sqrt{3} l^2 m n               P^{\parallel}_{pd\sigma}
+ m n \qty(1-2l^2)               P^{\parallel}_{pd\pi}
- \sqrt{3} l^2 m n				 P^{\perp}_{p\pi d\sigma}
- l^2 m n 		   			   P^{\perp}_{p\sigma d\pi}
- m n \qty(1-l^2) 		        P^{\perp}_{p\pi d\delta}$ \\
  \hline
\hline 
  $\mel{d_{yz}}{x}{d_{yz}}$ & $=$ &
  $0$ \\
   $\mel{d_{yz}}{y}{d_{yz}}$ & $=$ &
  $0$ \\
   $\mel{d_{yz}}{z}{d_{yz}}$ & $=$ &
  $0$ \\
  \hline
  $\mel{d_{yz}}{x}{d_{xz}}$ & $=$ &
  $
  \sqrt{3} m n^2                 P^{\perp}_{d\sigma d\pi}
 + m\qty(1-3n^2)	 		            P^{\perp}_{d\pi d\delta}$ \\
  $\mel{d_{yz}}{y}{d_{xz}}$ & $=$ &
  $
  -\sqrt{3} l n^2                 P^{\perp}_{d\sigma d\pi}
 - l\qty(1-3n^2)	 		            P^{\perp}_{d\pi d\delta}$ \\
   $\mel{d_{yz}}{z}{d_{xz}}$ & $=$ &
  $0$ \\
\hline
  $\mel{d_{yz}}{x}{d_{xy}}$ & $=$ &
  $
  \sqrt{3} m^2 n                 P^{\perp}_{d\sigma d\pi}
 + n\qty(1-3m^2)	 		            P^{\perp}_{d\pi d\delta}$ \\
   $\mel{d_{yz}}{y}{d_{xy}}$ & $=$ &
  $0$ \\
 
   $\mel{d_{yz}}{z}{d_{xy}}$ & $=$ &
  $
  -\sqrt{3} l m^2                 P^{\perp}_{d\sigma d\pi}
 - l\qty(1-3m^2)	 		            P^{\perp}_{d\pi d\delta}$ \\
  \hline
  \end{tabular}
  \label{tab:pol}
\end{table}
\twocolumngrid
\clearpage

\section{Full expressions of the coupling constants}
\label{sec:full_expr}

Here we give the full expressions calculated with perturbation theory of the coupling constants $\mathbb{A}_i$ and $\mathbb{B}_i$ calculated in Sec.~\ref{sec:mechanism}. The expressions are thus exact in $U$, $\lambda$, and $J_H$ in the limit of vanishing $\Delta$.

\subsection{Edge-sharing geometry}
\label{sec:expression_edge}
Except for $\mathbb{A}_1$, the nonvanishing coupling constants ($\mathbb{A}_2$, $\mathbb{A}_4$, $\mathbb{B}_{2-5}$) can be found from the operators $\hat M_u$, $\hat M_v$, and $\hat M_w$ defined in Eq.~\eqref{eq:matrices}. In the following, we give the explicit expression of $\mathbb{A}_1$ obtained from Eq.~\eqref{eq:pol3rd} (linear in $\Delta$), and the expressions for $\hat M_u$, $\hat M_v$, and $\hat M_w$ calculated with Eq.~\eqref{eq:Mpert} (here we only explicitly give the expressions without trigonal distortion, i.e., $\Delta=0$ and $t_4=0$).
As mentioned in Eq.~\eqref{eq:Aconstgeneral}, 
\begin{equation}
\mathbb{A}_1 = \sqrt{2}\frac{64}{81}\Delta (t_1 + t_2 - t_3 - t_4) \frac{J_H}{\lambda} \times \frac{P}{Q}.
\end{equation}
Without trigonal distortion, from Eq.~\eqref{eq:Mpert} we find $\hat M_{n, \textrm{eff}}  = \mathbb{M}_n \qty( \S_i \times \S_j) \cdot \vu n$ with
\begin{equation}
\mathbb{M}_n = -\frac{16}{9}\frac{R_n}{S},
\end{equation}
where $n=u,v,w$ (and $\vu n = \vu u, \vu v$, and $\vu w$, respectively).

The polynomials $P$, $Q$, $R_{u,v,w}$, and $S$ are given by
\begin{widetext}
\begin{align}
 P = & \qquad 2 (3 \lambda +U)^5 (3 \lambda +2 U) \Big(8 U^3 (t_1+t_2+t_3)+3 \lambda 
   U^2 (11 t_1+7 t_2+9 t_3)+9 \lambda ^3 (2 t_1+t_3)+24
   \lambda ^2 U (2 t_1+t_3)\Big) \nonumber \\
    & - J_H (3 \lambda +U)^2 \Big(448 U^6 (t_1+t_2+t_3)+112 \lambda  U^5
   (53 t_1+45 t_2+49 t_3)+2 \lambda ^2 U^4 (15659 t_1+10719
   t_2+13189 t_3) \nonumber \\
    &\qquad \qquad \qquad \qquad +18 \lambda ^3 U^3 (4717 t_1+2333 t_2+3525
   t_3)+72 \lambda ^4 U^2 (1732 t_1+498 t_2+1115 t_3) \nonumber \\
   & \qquad \qquad \qquad \qquad +27 \lambda ^5
   U (3446 t_1+336 t_2+1891 t_3)+13365 \lambda ^6 (2
   t_1+t_3)\Big) \nonumber \\
   & +3 J_H^2 (3 \lambda +U) \Big(9 \lambda ^6 (10474 t_1+672 t_2+5573
   t_3)+736 U^6 (t_1+t_2+t_3)+4 \lambda  U^5 (2923 t_1+2271
   t_2+2597 t_3) \nonumber \\
   & \qquad \qquad \qquad \qquad +8 \lambda ^2 U^4 (8882 t_1+5412 t_2+7147
   t_3)+10 \lambda ^3 U^3 (21673 t_1+9893 t_2+15783 t_3) \nonumber \\
   & \qquad \qquad \qquad \qquad +2 \lambda
   ^4 U^2 (176705 t_1+54033 t_2+115369 t_3)+15 \lambda ^5 U (19466
   t_1+3192 t_2+11329 t_3)\Big) \nonumber \\
   & - J_H^3 \Big(27 \lambda ^6 (54166 t_1+4640 t_2+29403 t_3)+3008 U^6
   (t_1+t_2+t_3)+48 \lambda  U^5 (1663 t_1+835 t_2+1249
   t_3) \nonumber \\
   & \qquad \qquad \qquad \qquad +300 \lambda ^2 U^4 (2109 t_1+697 t_2+1403 t_3)+2 \lambda ^3
   U^3 (1156429 t_1+268369 t_2+712399 t_3) \nonumber \\
   & \qquad \qquad \qquad \qquad +6 \lambda ^4 U^2 (723347
   t_1+116579 t_2+419963 t_3)+9 \lambda ^5 U (449042 t_1+48856
   t_2+248949 t_3)\Big) \nonumber \\
   &-  J_H^4 \Big(-17 \lambda ^5 (71174 t_1-9816 t_2+30679 t_3)+10784 U^5
   (t_1+t_2+t_3)+20 \lambda  U^4 (353 t_1+6237 t_2+3295
   t_3) \nonumber \\
   & \qquad \qquad \qquad \qquad +10 \lambda ^2 U^3 (-41005 t_1+54207 t_2+6601 t_3)-4 \lambda
   ^3 U^2 (423514 t_1-268196 t_2+77659 t_3) \nonumber \\
   & \qquad \qquad \qquad \qquad -14 \lambda ^4 U (176362
   t_1-63564 t_2+56399 t_3)\Big) \nonumber \\
   &- 6 J_H^5 \Big(2 \lambda ^4 (22394 t_1-45660 t_2-11633 t_3)-6944 U^4
   (t_1+t_2+t_3)-8 \lambda  U^3 (4555 t_1+8163 t_2+6359
   t_3) \nonumber \\
   & \qquad \qquad \qquad \qquad -3 \lambda ^2 U^2 (14845 t_1+70569 t_2+42707 t_3)+2 \lambda
   ^3 U (12386 t_1-132544 t_2-60079 t_3)\Big) \nonumber \\
   &- 36 J_H^6 \Big(4 \lambda ^3 (1238 t_1+560 t_2+899 t_3)+776 U^3
   (t_1+t_2+t_3)+3 \lambda  U^2 (1511 t_1+1707 t_2+1609
   t_3) \nonumber \\
   & \qquad \qquad \qquad \qquad +6 \lambda ^2 U (1342 t_1+1560 t_2+1451 t_3)\Big) \nonumber \\
   &- 216 J_H^7 \Big(\lambda ^2 (233 t_1+2637 t_2+1435 t_3)+344 U^2
   (t_1+t_2+t_3)+2 \lambda  U (395 t_1+999 t_2+697
   t_3)\Big) \nonumber \\
   &- 1296 J_H^8 (-\lambda  (137 t_1+315 t_2+226 t_3)-104 U
   + (t_1+t_2+t_3)) \nonumber \\
   &- 62208 J_H^9 (t_1+t_2+t_3),
\end{align}
\begin{align}
  Q &= (6 J_H-3 \lambda -2 U)^2 \left(6 J_H^2+J_H (4 \lambda +U)-U (3 \lambda
   +U)\right)^2 \left(6 J_H^2-J_H (17 \lambda +8 U)+(3 \lambda +U) (3 \lambda +2
   U)\right)^3
\end{align}
\begin{align}
  R_u = & (2 t_1+t_3) (3 \lambda +U)^2 (3 \lambda +2 U)^2 \nonumber \\
  & - 2 J_H (3 \lambda +U) \left(2 U^2 (13 t_1+t_2+7 t_3)+2 \lambda  U (53
   t_1+3 t_2+28 t_3)+51 \lambda ^2 (2 t_1+t_3)\right) \nonumber \\
  & + J_H^2 \left(\lambda ^2 (634 t_1+48 t_2+341 t_3)+8 U^2 (12 t_1+2
   t_2+7 t_3)+4 \lambda  U (126 t_1+16 t_2+71 t_3)\right) \nonumber \\
  & -12 J_H^3 ( \lambda (12 t_1 - 2t_2 + 5 t_3) + U(3t_1 - t_2 + t_3)) \nonumber \\
  & -36 J_H^4 (2 t_2+t_3),
\end{align}
\begin{align}
  R_v = & (2 t_1+t_3) (3 \lambda +U)^2 (3 \lambda +2 U)^2 \nonumber \\
  & - 2 J_H (3 \lambda +U) \left(2 U^2 (13 t_1-t_2+7 t_3)+2 \lambda  U (53
   t_1-3 t_2+28 t_3)+51 \lambda ^2 (2 t_1+t_3)\right) \nonumber \\
  & + J_H^2 \left(\lambda ^2 (634 t_1-48 t_2+341 t_3)+8 U^2 (12
   t_1-2 t_2+7 t_3)+4 \lambda  U (126 t_1-16 t_2+71
   t_3)\right) \nonumber \\
  & -12 J_H^3 (\lambda  (12 t_1+2 t_2+5 t_3)+U (3
   t_1+t_2+t_3)) \nonumber \\
  & -36 J_H^4 (t_3-2 t_2),
\end{align}
\begin{align}
  R_w = & (2 t_1+t_3) (3 \lambda +U)^2 (3 \lambda +2 U)^2 \nonumber \\
  & -2 J_H (3 \lambda +U) \left(51 \lambda ^2 (2 t_1+t_3)+4 U^2 (7 t_1+3
   t_3)+2 \lambda  U (56 t_1+25 t_3)\right) \nonumber \\
  & + J_H^2 \left(\lambda ^2 (682 t_1+293 t_3)+8 U^2 (14 t_1+5 t_3)+4
   \lambda  U (142 t_1+55 t_3)\right) \nonumber \\
  & -12 J_H^3 (\lambda  (10 t_1+7 t_3)+2 U (t_1+t_3)) \nonumber \\
  & - 36 J_H^4 (2 t_1-t_3),
\end{align}
\begin{align}
  S = (6 J_H-3 \lambda -2 U) \left(6 J_H^2+J_H (4 \lambda +U)-U (3 \lambda
   +U)\right) \left(6 J_H^2-J_H (17 \lambda +8 U)+(3 \lambda +U) (3 \lambda +2
   U)\right).
\end{align}
\end{widetext}

\subsection{Corner-sharing geometry}
\label{sec:expression_corner}
In the corner-sharing geometry, $\mathbb{A}_1 = \mathbb{A}_2 = 0$, and only the hopping polarization is nontrivial. It is calculated from the hopping operators $\hat N_y$ and $\hat N_z$ defined in Eq.~\eqref{eq:hop_N_op}. Without tetragonal distortion ($\Delta =0$), using Eq.~\eqref{eq:Mpert} we find $\hat N_{y, \textrm{eff}}= \mathbb{N} \qty( \S_i \times \S_j) \cdot \vu y$ and $\hat N_{z, \textrm{eff}} = \mathbb{N} \qty( \S_i \times \S_j) \cdot \vu z$, with $\mathbb{N} = \frac{32}{9} \times \frac{P'}{Q'}$ and
%\begin{align}
%  \mathbb{N} = \frac{8}{27} &t \Big(\frac{7 J_H-9 \lambda -3 U}{6 J_H^2-17 J_H \lambda -8 J_H U+9 \lambda ^2+2 U^2+9
%   \lambda  U} \nonumber \\
%   &+\frac{200 J_H^2}{(4 J_H-3 U) \left(6 J_H^2+J_H (4 \lambda +U)-U (3
%   \lambda +U)\right)} \nonumber \\
%   &+ \frac{3}{-6 J_H+3 \lambda +2 U}+\frac{36}{3 U-4 J_H} \Big).
%\end{align}
\begin{align}
  P'= & \left(9 \lambda ^2+2 U^2+9 \lambda  U\right)^2 \nonumber \\
  & -2 J_H (3 \lambda +U) \left(51 \lambda ^2+13 U^2+53 \lambda  U\right) \nonumber \\
  & + J_H^2 \left(317 \lambda ^2+48 U^2+252 \lambda  U\right) \nonumber \\
  & -18 J_H^3 (4 \lambda +U),
\end{align}
\begin{align}
  Q' = &(6 J_H-3 \lambda -2 U) \left(6 J_H^2+J_H (4 \lambda +U)-U (3 \lambda +U)\right) \nonumber \\
  &\times \left(6
   J_H^2-J_H (17 \lambda +8 U)+(3 \lambda +U) (3 \lambda +2 U)\right).
\end{align}

\clearpage
 
 \def\bibsection{\section*{\refname}} 
%\bibliographystyle{apsrev4-1}
%\bibliography{thesis.bib}

%
\end{document}